\documentclass[11pt]{article}   	
\usepackage{geometry}                		
\usepackage{graphicx}				
\usepackage{amssymb}
\usepackage{amsmath}
\usepackage{authblk}
\newcommand{\be}{\begin{equation}}
\newcommand{\ee}{\end{equation}}
\newcommand{\pom}{{I\!\!P}}
\newcommand{\regg}{{I\!\!R}}

\usepackage[utf8]{inputenc}
\usepackage{xspace}
\usepackage{lineno}

\usepackage[hypertexnames,
    pdftex,%
    colorlinks,%
		citecolor=blue,%
    hyperindex,%
    plainpages=false,%
    bookmarksopen,%
    bookmarksnumbered%
  ]{hyperref}
\usepackage[numbers,sort&compress]{natbib}

\newcommand\Eq[1]{(\ref{#1})}
\newcommand\Fig[1]{Fig.~\ref{#1}}

\DeclareRobustCommand\GeV{\ensuremath{\mathrm{GeV}}\xspace}
\DeclareRobustCommand\TeV{\ensuremath{\mathrm{TeV}}\xspace}

\DeclareRobustCommand\sred{\ensuremath{\sigma_\mathrm{r}}\xspace}

\DeclareRobustCommand\QQmin{\ensuremath{Q^2_\mathrm{min}}\xspace}
\DeclareRobustCommand\ximax{\ensuremath{\xi_\mathrm{max}}\xspace}

\newcommand\DD{\mathrm{D}}

\graphicspath{{figs/}}

\usepackage{color}
\usepackage[normalem]{ulem}
\definecolor{pncolor}{rgb}{0,0.1,0.7}
\definecolor{ascolor}{rgb}{1,0,1}
\definecolor{nacolor}{rgb}{1,0,0}
\definecolor{wscolor}{rgb}{0,0.6,0.2}

\DeclareRobustCommand\pnout{\bgroup\markoverwith{\color{pncolor}{\rule[0.4ex]{2pt}{0.8pt}}}\ULon}
\DeclareRobustCommand\asout{\bgroup\markoverwith{\color{ascolor}{\rule[0.4ex]{2pt}{0.8pt}}}\ULon}
\DeclareRobustCommand\naout{\bgroup\markoverwith{\color{nacolor}{\rule[0.4ex]{2pt}{0.8pt}}}\ULon}
\DeclareRobustCommand\wsout{\bgroup\markoverwith{\color{wscolor}{\rule[0.4ex]{2pt}{0.8pt}}}\ULon}

\title{\bf Inclusive diffraction in future   electron-proton and electron-ion colliders}
\author[1]{N\'estor Armesto}
\author[2]{Paul R. Newman}
\author[3]{Wojciech S\l{}omi\'nski}
\author[4]{ Anna M. Sta\'sto}
\affil[1]{\small \it Instituto Galego de F\'{\i}sica de Altas Enerx\'{\i}as IGFAE,
Universidade de Santiago de Compostela, 15782 Santiago de Compostela, Galicia-Spain}
\affil[2]{\small \it School of Physics and Astronomy, University of Birmingham, UK}
\affil[3]{\small \it Institute of Physics, Jagiellonian University, Krakow, Poland}
\affil[4]{\small \it Department of Physics, Penn State University, University Park, PA 16802, USA}

\begin{document}
\maketitle
\begin{abstract}
We analyse the possibilities for the study of inclusive diffraction offered by 
future electron--proton/\-nu\-cleus colliders in the TeV regime, the Large Hadron-electron Collider as an upgrade of the
 HL-LHC and the Future Circular Collider in electron-hadron mode. Compared to $ep$ collisions at HERA,
  we find an extension  of the  available kinematic range in $x$ by a factor of order $20$ and of the maximum $Q^2$ by 
a factor of order $100$ for LHeC, while the FCC version would extend 
the coverage by a further order of magnitude both in $x$ and $Q^2$.  
This translates into a range of available  momentum fraction of the 
diffractive exchange with respect to the hadron ($\xi$), 
down to $10^{-4}-10^{-5}$ for a wide range of the momentum fraction 
of the parton with respect to the diffractive exchange ($\beta$). 
Using the same framework and methodology employed in previous studies at 
HERA, considering only the experimental uncertainties and not those stemming from the functional form of the initial conditions or other ones of theoretical origin, and  under very conservative assumptions for the luminosities and 
systematic errors, we find an improvement in the extraction of 
diffractive parton densities from fits to reduced cross sections 
for inclusive coherent diffraction in $ep$ by 
about an order of magnitude.
For $eA$, we also perform the simulations for the Electron Ion Collider. We find that an extraction of the currently unmeasured nuclear diffractive parton densities is possible with similar accuracy to that in $ep$.
\end{abstract}

\section{Introduction}

Deep Inelastic Scattering (DIS) of a lepton on a proton is the cleanest 
way to explore the proton structure. The HERA accelerator at Hamburg was the 
only  $ep$ collider to date. It scattered electrons and positrons on protons, 
at  centre-of-mass energy $\sqrt{s} = 318\, \GeV$.  One of the most 
striking discoveries at HERA was the observation of the strong rise of the 
gluon density at small values of Bjorken $x$.  HERA provided the measurement 
of the parton densities with high accuracy, necessary for precise theoretical 
calculations of 
a vast range of processes under study
at the Large Hadron Collider (LHC). Another discovery of HERA was the observation of 
a large ($\sim 10\%$) fraction of diffractive events 
in DIS \cite{Adloff:1997sc,Breitweg:1997aa}, see the review \cite{Newman:2013ada} and refs. therein. In these events the proton 
stays intact or dissociates into a state with the proton quantum numbers, 
despite undergoing a violent, highly energetic collision, 
and is separated from the rest of the produced particles by a large rapidity gap.  
In a series of ground-breaking papers, the HERA experiments determined the deep inelastic
structure of the $t$-channel exchange in these events in the form of diffractive parton
densities.

The precise measurement of diffraction in DIS is of great importance for our understanding of the strong interaction. 
First, the mechanism through which a composite strongly interacting object interacts perturbatively while 
keeping colour neutrality offers information about the confinement mechanism. 
Second, diffraction is known to be highly sensitive to the low-$x$ partonic content of the proton 
and its evolution with energy and it therefore has considerable promise to reveal deviations 
from standard linear evolution through higher twist effects or, eventually, non-linear dynamics. 
Third, it allows checks of basic theory predictions such as the relation between diffraction 
in $ep$ scattering and nuclear shadowing  \cite{Gribov:1968jf}. Finally, the accurate 
extraction of diffractive parton distribution functions facilitates tests of the  
range of validity of the perturbative factorization \cite{Collins:1997sr,Berera:1995fj,Trentadue:1993ka}. We note that our study is only about inclusive diffraction. Other aspects could be addressed, for example factorization breaking in diffractive dijet photoproduction \cite{Klasen:2008ah,Frankfurt:2011cs} which can also be studied in ultraperipheral collisions at the LHC \cite{Guzey:2016tek}.

Diffraction has also been studied outside the collinear framework, for example in the Color Glass Condensate \cite{Gelis:2010nm} and its implementation in photoproduction and DIS through the dipole model \cite{Nikolaev:1990ja,Mueller:1993rr}. Differences are expected with respect to collinear factorization at small $x$, resulting in particularly strong  modifications  of diffraction in nuclei with respect to proton \cite{Nikolaev:1995xu,Frankfurt:1991nx}, see detailed discussions in \cite{Kowalski:2008sa}.

Future DIS machines could explore this phenomenon at higher energies 
and with much  higher precision. The Large Hadron-electron Collider (LHeC) 
is a proposal \cite{Dainton:2006wd, AbelleiraFernandez:2012cc, Klein:2018rhq} 
for an $ep$ and $eA$ machine at CERN. It would 
utilize the $7 \, \TeV$ proton beam from the LHC and collide it 
with a $60\, \GeV$ electron beam accelerated by an energy recovery  linac, 
thus reaching a centre-of-mass energy $\sqrt{s} = 1.3 \, \TeV$.  
Dedicated studies of the machine parameters  \cite{Bordry:2018gri, LHeClumi}  
show that  its peak luminosity would reach  $10^{34}\, {\rm cm}^{-2} {\rm s}^{-1}$, 
about three orders of magnitude higher than HERA. The projected running of the machine is over three periods. In the initial run period the total integrated luminosity is estimated to be 
$50 \; {\rm fb}^{-1}$. Throughout the 
entire  operation the LHeC is projected to reach $1 \,{\rm ab}^{-1}$ integrated luminosity.  
It would also be the first electron--nucleus collider, as it would 
scatter electrons on a beam of  nuclei from the LHC, with an energy of $2.75 \, \TeV$ per nucleon resulting  in the centre-of-mass energy per nucleon $\sqrt{s}=812 \, \GeV$. The integrated luminosity for  collisions on nuclei is projected to be 
of the order $10 \, {\rm fb}^{-1}$ which is ten times larger than the total luminosity 
collected in $ep$ at HERA.   This would allow 
measurements of nuclear structure with unprecedented precision.  
Beyond LHeC, the next generation $ep$ collider would be the Future Circular 
Collider in electron-hadron mode (FCC-eh), utilizing the $50 \, \TeV$ proton beam 
from the FCC \cite{FCC_CDRv1,FCC_CDRv3} which would 
probe DIS at centre-of-mass energy of $\sqrt{s} = 3.5 \, \TeV$ 
with a total integrated luminosity of several ${\rm ab}^{-1}$. The $eA$ collisions at the FCC-eh \cite{Bordry:2018gri, LHeClumi} would be performed with a lead beam with energy per nucleon  $19.7 \, \TeV$ which would give a  centre-of-mass energy per nucleon of $\sqrt{s}=2.2 \, \TeV$. At lower energies $\sqrt{s} \sim 0.1\, \TeV$, the Electron Ion Collider (EIC) \cite{Accardi:2012qut} will also measure diffraction, covering a smaller kinematic region than HERA in $ep$ but a completely novel region in $e$A with respect to fixed target experiments where diffraction has been barely studied.

These machines would facilitate the 
study the proton and nuclear structure with  extremely high precision. They would  unravel complete details of the partonic structure of the proton,  explore novel QCD  dynamics at small values of Bjorken $x$, constrain the Higgs properties, perform searches for physics beyond the Standard Model, and provide 
complementary precision measurements of electroweak physics to 
$e^+e^-$ colliders and the LHC and FCC-hh. DIS on nuclei 
would allow 
the study of nuclear structure in a previously unexplored kinematic 
region in $(x,Q^2)$. It is therefore  expected to thoroughly transform our present knowledge on
 parton structure in nuclei,  also largely strengthening  the chromodynamic base for the Quark Gluon Plasma and the ridge correlation
 phenomenon.
 
In this work we perform a thorough analysis of the capability of the 
LHeC and FCC-eh machines to explore inclusive diffraction in DIS. 
We first determine the accessible kinematic range for diffraction  of both machines. 
Using a very conservative assumption of $2 \, {\rm fb}^{-1}$ for the 
integrated luminosity we perform a 
simulation of the data for 
inclusive coherent $ep$ diffraction in the projected parameter space. 
This is performed by extending the fits 
used to extract the diffractive parton densities (DPDFs) at HERA. 
We then demonstrate the potential of both machines to constrain the  
DPDFs and point out the sensitivity to the interesting region of low $Q^2$ where deviations from standard linear evolution could become sizeable.  
These machines would also be able to explore the top quark contribution 
to diffraction as well as measuring diffraction in the 
charged current exchange, though we do not perform analysis of these interesting phenomena here.
We analyse the sensitivity to kinematic cuts and variations of the fit framework.  
We also note the possibilities that measurements at these new machines offer to improve 
existing constraints on the shape of the gluon distribution, and
the sensitivity to physics beyond linear twist-2 DGLAP evolution at moderate $Q^2$.

We also perform a simulation of the diffractive pseudodata for  $eA$ collisions for different scenarios of nuclear shadowing. Nuclear diffractive parton distributions have never been measured and therefore 
the considered machines would be the first to extract these important quantities. It would also be possible to investigate the relation between nuclear shadowing and diffraction.
 
We focus on the impact of the new kinematic range and expected increase in the measurement accuracy on the extraction of diffractive parton densities from fits to cross sections for inclusive coherent diffraction in $ep$.
To this end we
stick to the
parametrization model used in the HERA fits
and we estimate the experimental uncertainties of DPDFs obtained from fits to pseudo-data generated for the LHeC/FCCeh scenario.
Comparing these uncertainties to the ones resulting from the HERA data
we observe an order of magnitude improvement.
It should be noted that the values of these DPDF uncertainties come from the {\it experimental uncertainties only}, and as such they are most probably below the expected full uncertainties which would also arise from the parametrization and theory uncertainties.
Nevertheless the relative improvement clearly shows the measurement potential of the new machines.

The structure of the paper is the following. In Sec.~\ref{sec:sec2} we recall  the 
formulae for the diffractive cross sections, the factorization of the inclusive 
diffractive structure functions and the origin of their sensitivity to DPDFs. 
In Sec.~\ref{sec:sec3} we present the details of the simulations for the diffractive DIS. In particular, in subsection ~\ref{sec:dpdf_param} we discuss the parametrization used  at HERA, in \ref{sec:kinematics} we show the details of the diffractive kinematic range in new machines, and in \ref{sec:pseudo_data}  the method to obtain the projected pseudodata with errors is discussed. In Sec.~\ref{sec:res} we present our fitting methodology and the potential for  constraining 
the diffractive parton densities by both machines. Sec.~\ref{sec:nuclei} is devoted to the prospects of the diffractive deep inelastic in nuclei.
Finally we summarize our findings in Sec.~\ref{sec:conclusions}.

\section{Diffractive cross section and diffractive PDFs}
\label{sec:sec2}

\begin{figure}
\centering{\includegraphics[height=5cm]{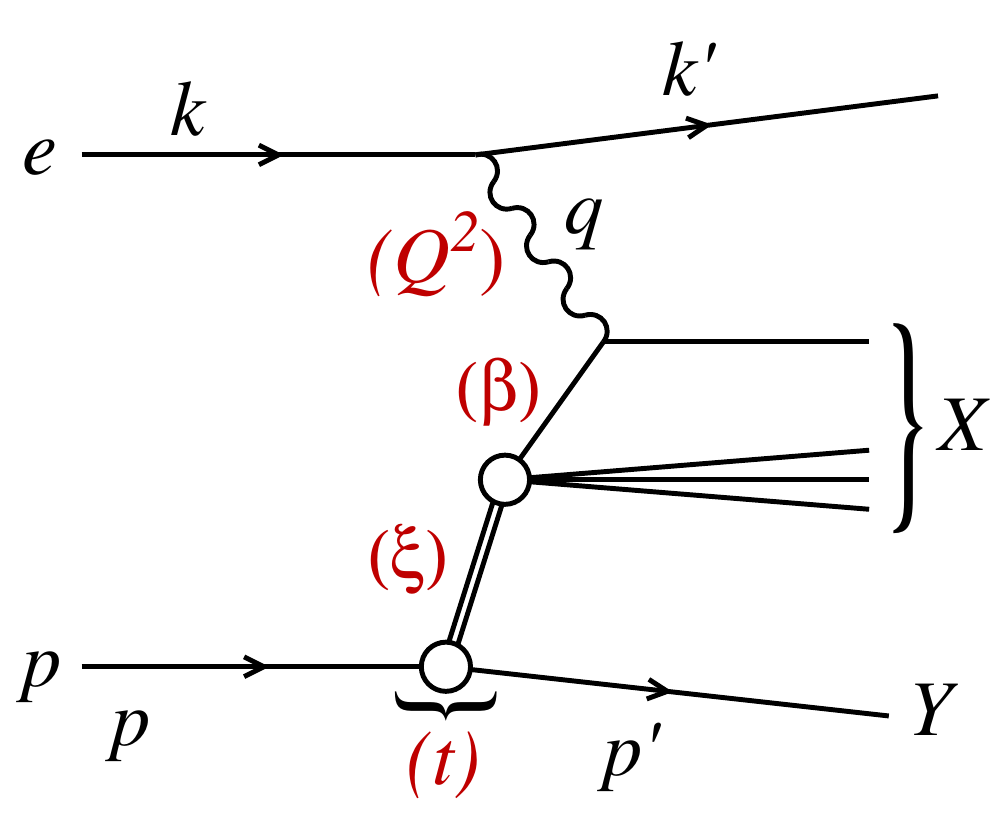}}
\caption{A diagram of a diffractive NC event in DIS together with the corresponding variables, in the one-photon exchange approximation. The large rapidity gap is between the system $X$ and the scattered proton $Y$ (or its low mass excitation).}
\label{fig:ddis}
\end{figure}

In \Fig{fig:ddis} we show a diagram depicting a neutral current diffractive deep inelastic event.
 Charged currents could also be considered and they were measured at HERA \cite{Aktas:2006hy} but with large statistical uncertainties and in a very restricted region of phase space. Although they could be measured at both the LHeC and the FCC-eh with larger statistics and more extended kinematics, in this first study we limit ourselves to neutral currents.
The incoming electron or positron, with four momentum $k$, scatters off the proton, with incoming momentum $p$, and the interaction proceeds through the exchange of a virtual photon with four-momentum $q$. The kinematic variables   
for such an event include the standard deep inelastic variables
 \be 
 Q^2=-q^2\,, \qquad  x=\frac{-q^2}{2p\cdot q}\,,  \qquad y=\frac{p\cdot q}{p\cdot k}\,, 
 \ee
 where $Q^2$ describes the photon virtuality, $x$ is the Bjorken variable and $y$ the inelasticity of the process. In addition, the variables
 \be
 s=(k+p)^2\,, \qquad  W^2=(q+p)^2 \, ,
 \ee
 are  
 the electron-proton centre-of-mass energy squared and  the photon-proton centre-of-mass energy squared, respectively. The distinguishing feature of the diffractive event ${ep\rightarrow eXY}$ is  the presence of the large rapidity gap between the diffractive system, characterized by the invariant mass 
$M_X$ and the final proton (or its low-mass excitation) $Y$
 with four momentum $p'$. 
In addition to the standard DIS variables listed above, diffractive events are also characterized by an additional set of variables defined as
\be
t=(p-p')^2\,, \qquad \xi=\frac{Q^2+M_X^2-t}{Q^2+W^2}\,, \qquad \beta = \frac{Q^2}{Q^2+M_X^2-t}\, .
\ee
In the above  $t$ is the squared
 four-momentum transfer   at the proton vertex, $\xi$ (alternatively denoted by $x_\pom$)  can be interpreted as  the momentum fraction of the `diffractive exchange'   with respect to the hadron,  and  $\beta$ 
is the momentum fraction of the parton with respect to the diffractive exchange. 
 The two momentum fractions combine to give  Bjorken-$x$, $x=\beta \xi$.
 
The physical picture suggested by \Fig{fig:ddis} is that the initial proton splits into a final state $Y$ of momentum $p' \simeq (1-\xi)p$ and the object which is responsible for the diffractive exchange of momentum $\xi p$. The latter 
in turn undergoes a DIS-like process to produce the final state $X$ (see Sec.~\ref{sec:dpdf_param} for more details).
The study presented in this paper concerns coherent diffraction 
(i.e. the non-dissociating case), where the final state $Y$ is a proton.
Experimentally, this requires tagging of the final proton,
which was performed at HERA using Roman pot insertions to the forward beam-pipe, for example the FPS (LPS) of the H1 (ZEUS) collaborations. 
Most of the HERA data are based, however, on the large rapidity gap (LRG) 
technique, which results in a small proton dissociative admixture -- the response from 
detector components at very forward rapidities, supplemented with dedicated  MC modelling, were used to normalize these results to the coherent cross-sections \cite{Aktas:2006hy,Chekanov:2008fh}.

Diffractive cross sections
 in the neutral current case can be presented in the form of the  reduced cross 
sections \cite{Aktas:2006hy}
\begin{subequations}
\be
\frac{d^4 \sigma^{\DD}}{d\xi d\beta dQ^2 dt} = \frac{2\pi \alpha_{\rm em}^2}{\beta Q^4} \, Y_+ \, \sred^{\DD(4)}\,  ,
\label{eq:sigmared4}
\ee
or, upon integration over $t$,
\be
\label{eq:sigmared3}
\frac{d^3 \sigma^{\DD}}{d\xi d\beta dQ^2} = \frac{2\pi \alpha_{\rm em}^2}{\beta Q^4} \, Y_+ \, \sred^{\DD(3)}\, , 
\ee
\end{subequations}
where
$Y_+= 1+(1-y)^2$ and the reduced cross sections can be expressed in terms of two diffractive structure functions
$F_2^{\DD}$ and $F_\mathrm{L}^{\DD}$. 
In the one-photon approximation, the relations are
\begin{subequations}
\be
\label{eq:sred3}
\sred^{\DD(3)} = F_2^{\DD(3)}(\beta,\xi,Q^2) - \frac{y^2}{Y_+} F_\mathrm{L}^{\DD(3)}(\beta,\xi,Q^2) \; ,
\ee
\be
\sred^{\DD(4)} = F_2^{\DD(4)}(\beta,\xi,Q^2,t) - \frac{y^2}{Y_+} F_\mathrm{L}^{\DD(4)}(\beta,\xi,Q^2,t) \; .
\ee
\end{subequations}
Note that the structure functions $F_\mathrm{2,L}^{\DD(4)}$ have dimension $\GeV^{-2}$, while $F_\mathrm{2,L}^{\DD(3)}$ are dimensionless.
In this analysis we neglect $Z^0$ exchange, though it should be included in 
future studies.

The reduced cross sections $\sred^\DD$ depend on  centre-of-mass 
energy via $y = \frac{Q^2}{\xi\beta s}$. The $Y_+$ factors ensure that
in the region where $y$ is not too close to unity, 
\be
\sred^{\DD} \simeq F_2^{\DD} 
\ee
to good approximation. 

Both $\sred^{\DD(3)}$ and $\sred^{\DD(4)}$ have been measured at the HERA collider  \cite{Adloff:1997sc,Breitweg:1997aa,Chekanov:2005vv, Aktas:2006hx, Aktas:2006hy,Chekanov:2008fh,Chekanov:2009aa,Aaron:2010aa,Aaron:2012ad} and used to obtain QCD-inspired parametrizations.

The standard perturbative QCD approach to diffractive cross sections is based on collinear factorization \cite{Collins:1997sr,Berera:1995fj,Trentadue:1993ka}. It was demonstrated that, similarly to the inclusive DIS cross section, the diffractive cross section can be written, up to terms of  order  ${\cal O}(1/Q^2)$, in a factorized form
\be
d\sigma^{ep\rightarrow eXY}(\beta,\xi,Q^2,t) \; = \; \sum_i \int_{\beta}^{1} dz \ d\hat{\sigma}^{ei}\left(\frac{\beta}{z},Q^2\right) \, f_i^{\rm D}(z,\xi,Q^2,t) \; ,
\label{eq:collfac}
\ee
where the sum is performed over all parton flavours (gluon, $d$-quark, $u$-quark, etc.).
The hard scattering partonic cross section $d\hat{\sigma}^{ei}$ can be computed perturbatively in QCD and is the same as in the inclusive deep inelastic scattering case. The long distance 
part $f_i^{\rm D}$ corresponds to the diffractive parton distribution functions,
which can be interpreted as conditional probabilities for partons in the proton, provided the proton is scattered into the final state system $Y$ with specified 4-momentum $p'$. 
They are evolved using the DGLAP evolution equations \cite{Gribov:1972rt,Gribov:1972ri,Altarelli:1977zs,Dokshitzer:1977sg} similarly to the inclusive case.
The analogous formula for the $t$-integrated structure functions reads
\begin{equation}
\label{eq:FD3-fac}
F_{2/\mathrm{L}}^{\DD(3)}(\beta,\xi, Q^2) =
\sum_i \int_{\beta}^1 \frac{dz}{z}\,
	 C_{2/\mathrm{L},i}\Big(\frac{\beta}{z}\Big)\, f_i^{\DD(3)}(z,\xi,Q^2) \; ,
\end{equation}
where the coefficient functions $C_{2/\mathrm{L},i}$ are the same as in  inclusive DIS.

\section{Simulations for the electron-proton DIS}
\label{sec:sec3}

\subsection{Diffractive PDF parametrizations and HERA data}
\label{sec:dpdf_param}

Fits to the diffractive structure functions were performed by  
H1 \cite{Aktas:2006hy} and ZEUS \cite{Chekanov:2009aa}. They both 
parametrize the diffractive PDFs in a two component 
model, which is a sum of two exchange contributions, $\pom$ and $\regg$:
\be
f_i^{\DD(4)}(z,\xi,Q^2,t) =  f^p_{\pom}(\xi,t) \, f_i^{\pom}(z,Q^2)+f^p_{\regg}(\xi,t) \, f_i^{\regg}(z,Q^2) \;.
\label{eq:param_2comp}
\ee
For both of these terms proton vertex factorization is assumed, meaning that the diffractive exchange can be interpreted as colourless objects called a `Pomeron' or a `Reggeon' with  parton distributions $f_i^{\pom,\regg}(\beta,Q^2)$. The flux factors  $f^p_{\pom,\regg}(\xi,t)$ represent the probability that a Pomeron/Reggeon with given values $\xi,t$ couples to the proton.  They are parametrized using the form motivated by  Regge theory,
\be
 f^p_{\pom,\regg}(\xi,t) = A_{\pom,\regg} \frac{e^{B_{\pom,\regg}t}}{\xi^{2\alpha_{\pom,\regg}(t)-1}} \; ,
\label{eq:flux}
\ee
with a linear trajectory ${\alpha_{\pom,\regg}(t)=\alpha_{\pom,\regg}(0)+\alpha_{\pom,\regg}'\,t}$.
The diffractive PDFs relevant to the $t$-integrated cross-sections read
\be
f_i^{\DD(3)}(z,\xi,Q^2) =  \phi^{\;p}_{\pom}(\xi) \, f_i^{\pom}(z,Q^2) + \phi^{\;p}_{\regg}(\xi) \, f_i^{\regg}(z,Q^2) \; ,
\label{eq:fD3_2comp}
\ee
with
\begin{equation}
	 \phi^{\;p}_{\pom,\regg}(\xi) = \int \! dt\; f^p_{\pom,\regg}(\xi,t) \;.
\end{equation}
Note that, the notions of `Pomeron' and `Reggeon' used here to model 
hard diffraction in DIS are, in principle, different from those describing the soft hadron-hadron interactions; in particular, the parameters of the fluxes may be different.

The diffractive parton distributions of the Pomeron at the initial scale $\mu_0^2 = 1.8\,\GeV^2$ are parametrized as
\be
z f_i^\pom (z,\mu_0^2)= A_i z^{B_i} (1-z)^{C_i} \; ,
\label{eq:initcond}
\ee
where $i$ is a gluon or a light quark. In the diffractive parametrizations all the 
light quarks (anti-quarks) are assumed to be equal. 
For the treatment of heavy flavours, a variable flavour number  scheme (VFNS) is adopted, where the charm and bottom quark DPDFs are generated radiatively via DGLAP evolution,  and no intrinsic heavy quark distributions are assumed.
The structure functions are calculated in a General-Mass Variable Flavour Number  scheme (GM-VFNS) \cite{Collins:1986mp,Thorne:2008xf} which 
ensures a smooth transition of $F_\mathrm{2,L}$ across the flavour thresholds by including
$\mathcal{O}(m_h^2/Q^2)$ corrections.
The parton distributions for the Reggeon component are taken from a parametrization which was obtained from fits to the pion structure function \cite{Owens:1984zj,Gluck:1991ey}.

In Eq. \Eq{eq:param_2comp} the normalization factors of fluxes, $A_{\pom,\regg}$ and of DPDFs, $A_i$ enter in the product. To resolve the ambiguity we fix\footnote{Here, as in the HERA fits, $A_{\pom}$ is fixed by normalizing $\phi^{\;p}_{\pom}(0.003) = 1$.} 
$A_{\pom}$
and use $f_i^{\regg}(z,Q^2)$ normalized to the pion structure function,
which results in $A_i$ and $A_{\regg}$ being well defined free fit parameters.

\begin{figure}
\centerline{%
	\includegraphics*[width=0.4\textwidth,trim=0 0 0 50]{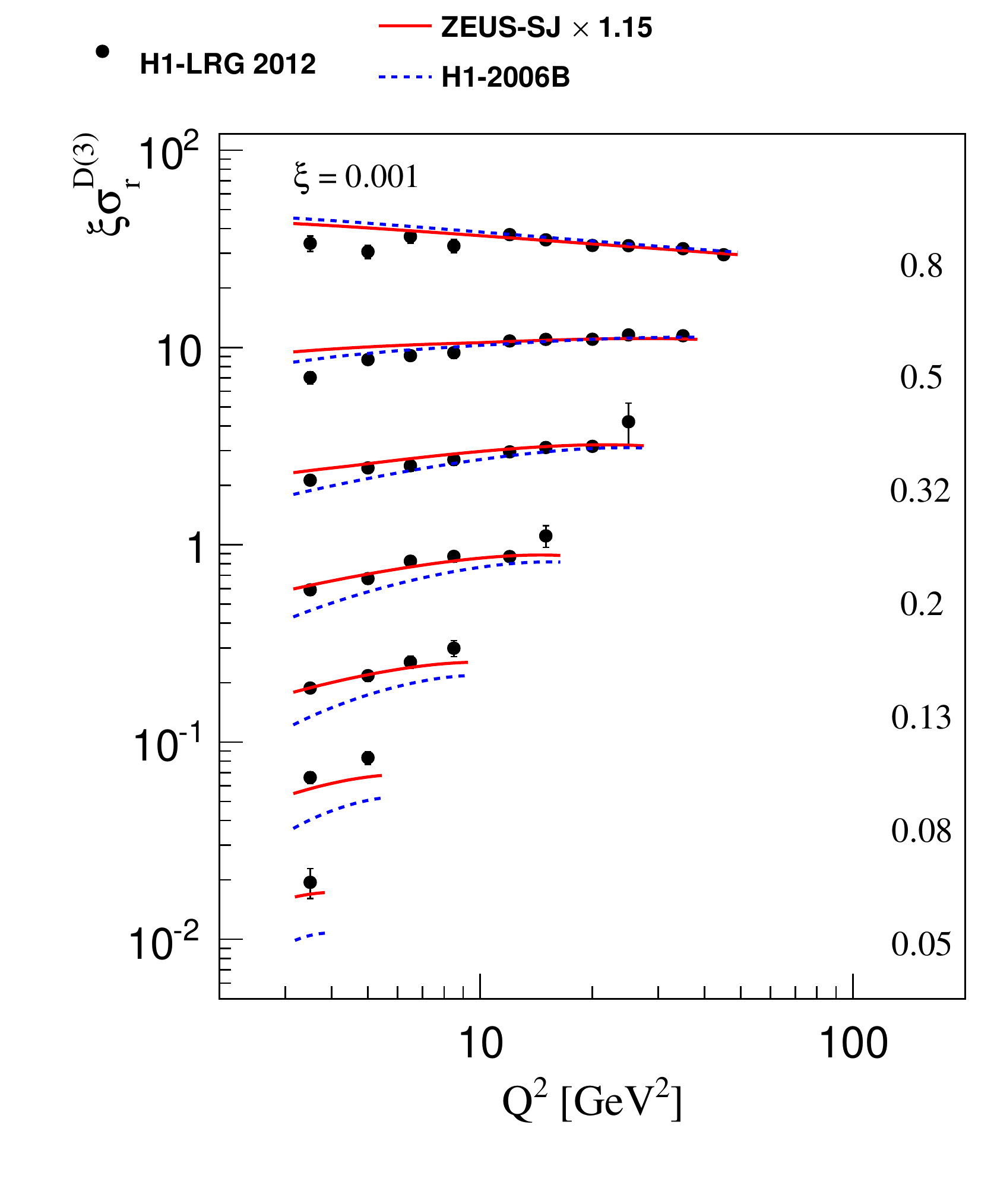}%
	\includegraphics*[width=0.4\textwidth,trim=0 0 0 50]{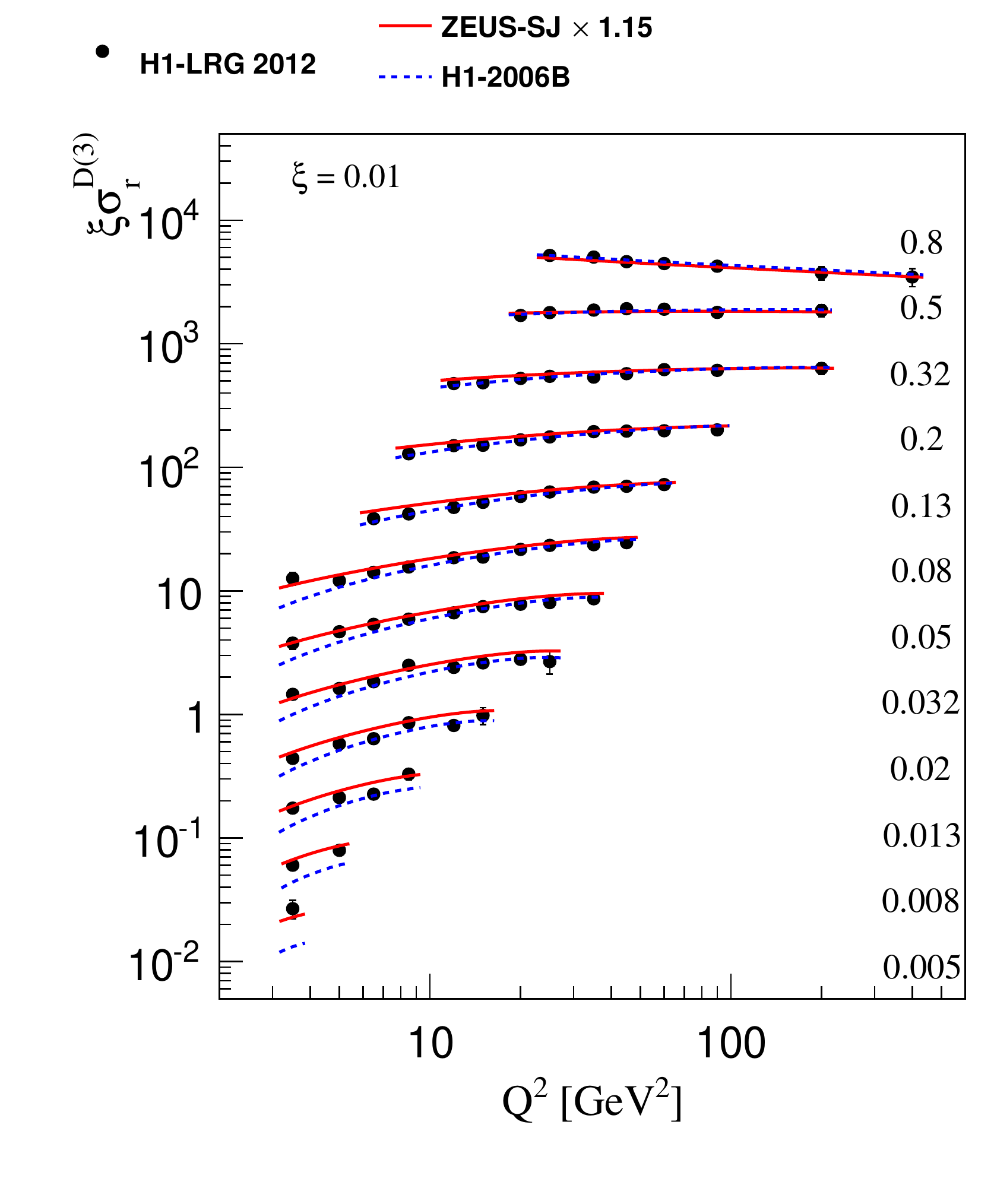}%
}
\caption{Experimental data from the H1 collaboration at HERA \cite{Aaron:2012ad} 
on the reduced diffractive cross section as a function of $Q^2$ in bins of 
$\beta$ for two values of $\xi=0.001$ (left)  and $\xi=0.01$ (right). 
The lines indicate predictions from two fits to older data: H1 2006 Fit B (dotted, blue) and ZEUS-SJ (solid, red). The values shown are scaled by $3^k$ for $k = 0,1,\dots$ upwards.}
\label{fig:herafits}
\end{figure}


There are different types of diffractive fits available in the literature. Here we mention the NLO parametrizations from HERA relevant to the current study:
\begin{description}
\item[Fit-S:] All parameters $A_i,B_i,C_i$ are free, as well as $A_{\regg}$ and $\alpha_{\pom,\regg}(0)$ (9 parameters). This is the ZEUS-S fit.
\item[Fit-C:] Parameters $B_g,C_g$ are set to zero, resulting  in 
a constant gluon density at the starting scale for QCD evolution. 
This corresponds to the `H1 Fit B' fit when $A_{\regg}$ and $\alpha_{\pom}(0)$ are free 
(6 parameters), and to the ZEUS-C fit when $A_{\regg}$ and $\alpha_{\pom,\regg}(0)$
are free (7 parameters).
\item[Fit-SJ:] All parameters $A_i,B_i,C_i$ are free. In addition,   
dijet production data are used to constrain the gluon. 
This amounts to the ZEUS-SJ fit when $A_{\regg}$ and $\alpha_{\pom,\regg}(0)$
are free (9 parameters) and to the H1-2007 fit \cite{Aktas:2007bv} 
when $A_{\regg}$ and $\alpha_{\pom}(0)$ are free (8 parameters).
\end{description}

Note that Fit-S and Fit-SJ share the same functional form, differing only in the use of  dijet data in the latter.
In the current work the ZEUS-SJ fit predictions are used for pseudodata simulation
and the fitting analysis is performed with the Fit-S parametrization model, i.e. using the same parametrization.

In \Fig{fig:herafits} we show some example of recent HERA data \cite{Aaron:2012ad} 
compared with two fits, H1 Fit B and ZEUS-SJ. 
Note that the fits were performed to  older data than shown in the Figure.

\subsection{LHeC and FCC-eh kinematics compared with HERA data}
\label{sec:kinematics}
\begin{figure}
\centering{\includegraphics[width=10cm]{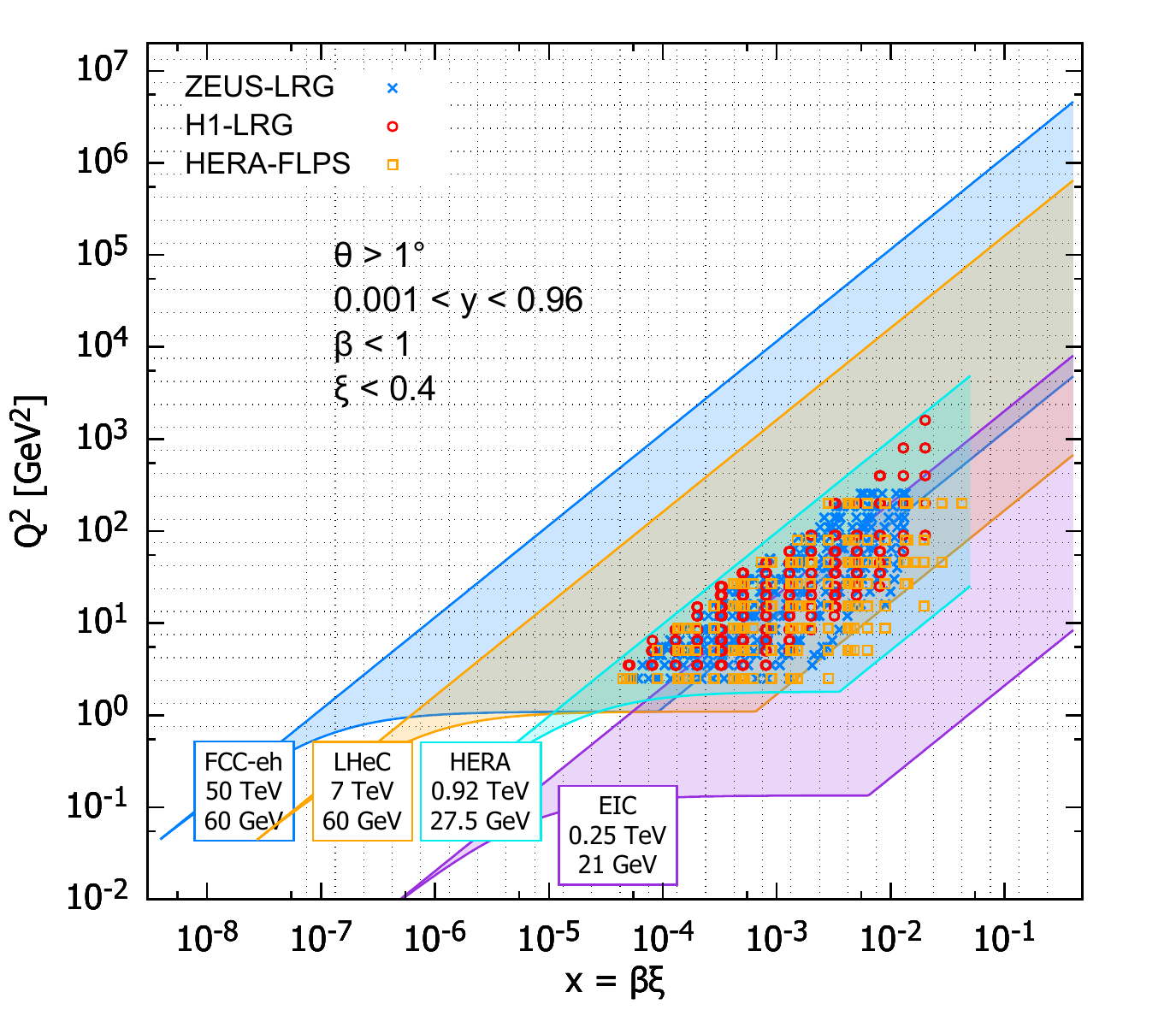}}
\caption{Kinematic phase space for inclusive diffraction in $(x,Q^2)$ for  the EIC (magenta region), the LHeC (orange region) and the FCC-eh (dark blue region) as compared with the HERA data (light blue region, ZEUS-LRG~\cite{Chekanov:2008fh}, H1-LRG~\cite{Aaron:2012ad}, HERA-FLPS~\cite{Aaron:2012hua}). The acceptance limit for the electron in the detector design has been assumed to be $ 1^{\circ}$, and we take $\xi<0.4$.}
\label{fig:phasespace_xQ}
\end{figure}

The  kinematic range in $(\beta,Q^2,\xi)$ is restricted by the following cuts:
\begin{itemize}
\item $Q^2 \ge 1.8\,\GeV^2$: due to the fact that the initial distribution for the DGLAP evolution is parametrized at $\mu_0^2=1.8 \,\GeV^2$. The renormalization and factorization scales are taken to be equal to $Q^2$.
\item $\xi<0.4$ : by physical and experimental limitations. This rather high $\xi$ value is an experimental challenge and physically enters the phase-space region where the Pomeron 
contribution should become negligible. 
Within the two-component model of Eq. \Eq{eq:param_2comp}, at high $\xi$ the cross-section is dominated by the secondary Reggeon contribution, which is poorly fixed by the HERA data. We present this high $\xi$ ($> 0.1$) region for illustrative purpose and for the sake of discussion of the fit results in Sec.~\ref{sec:res}.
\end{itemize}

In \Fig{fig:phasespace_xQ} the accessible kinematic range in $(x,Q^2)$ is 
shown for three machines: HERA, LHeC and FCC-eh. For the LHeC design the 
range in $x$ is increased by a factor $\sim 20$ over HERA
and the maximum available $Q^2$ by a factor $\sim 100$. The FCC-eh machine would further increase this range with respect to LHeC by roughly one order of magnitude in both $x$ and $Q^2$. We also show the EIC kinematic region for comparison.

\begin{figure}
\centering{\includegraphics*[width=12cm,trim=0 0 0 50]{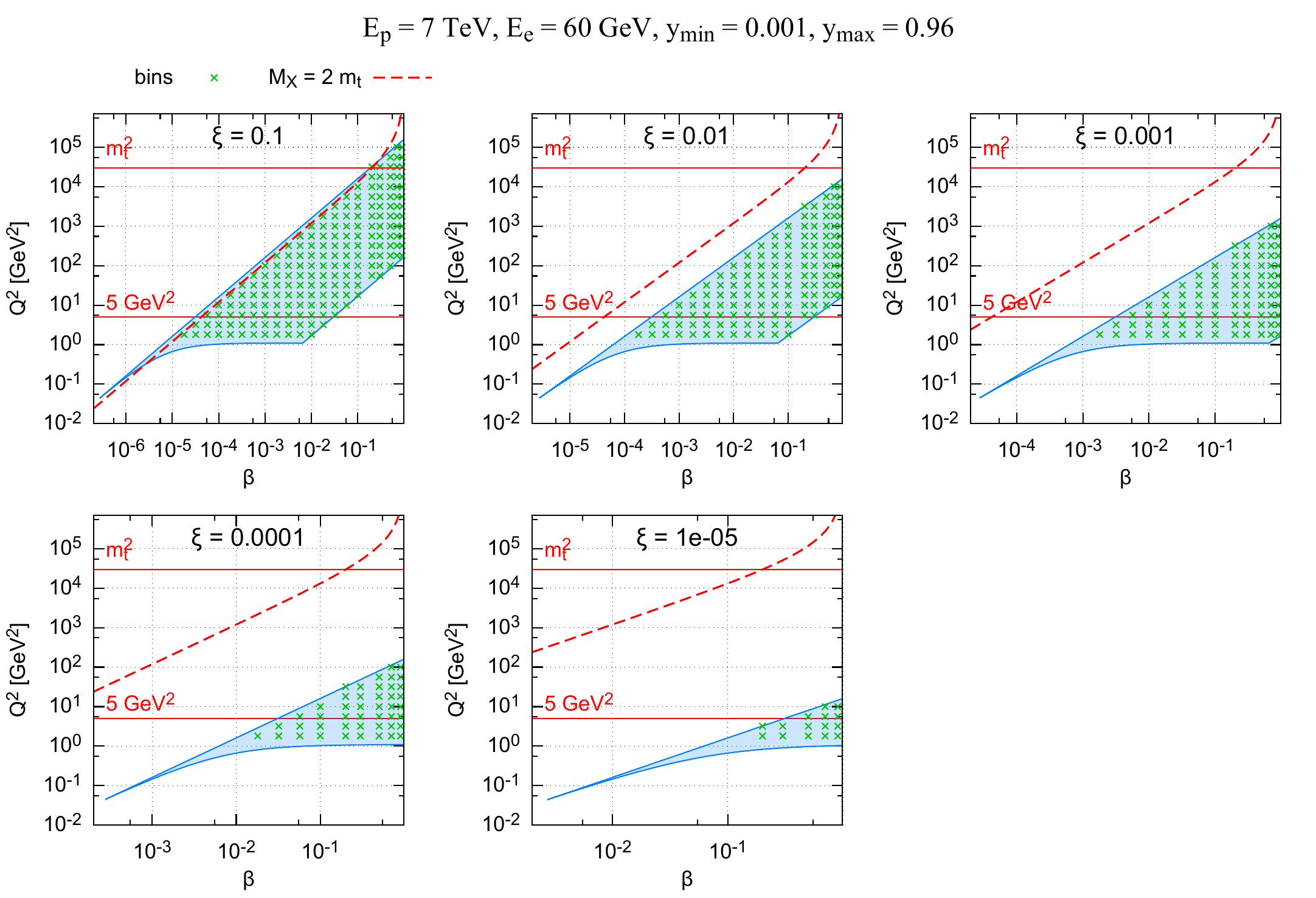}}
\caption{Kinematic phase space for inclusive diffraction in $(\beta,Q^2)$ for fixed values of $\xi$ for the LHeC design.
The horizontal lines indicate correspondingly, $Q_{\rm}^2=5 \; {\rm GeV}^2$, the lowest data value for the DGLAP fit performed in this study and $m_{\rm t}^2$ the 6-flavour threshold. The dashed line marks the kinematic limit for $t\bar t$ production.
}
\label{fig:phasespace_bQ_lhec}
\end{figure}

\begin{figure}
\centering{\includegraphics*[width=12cm,trim=0 0 0 50]{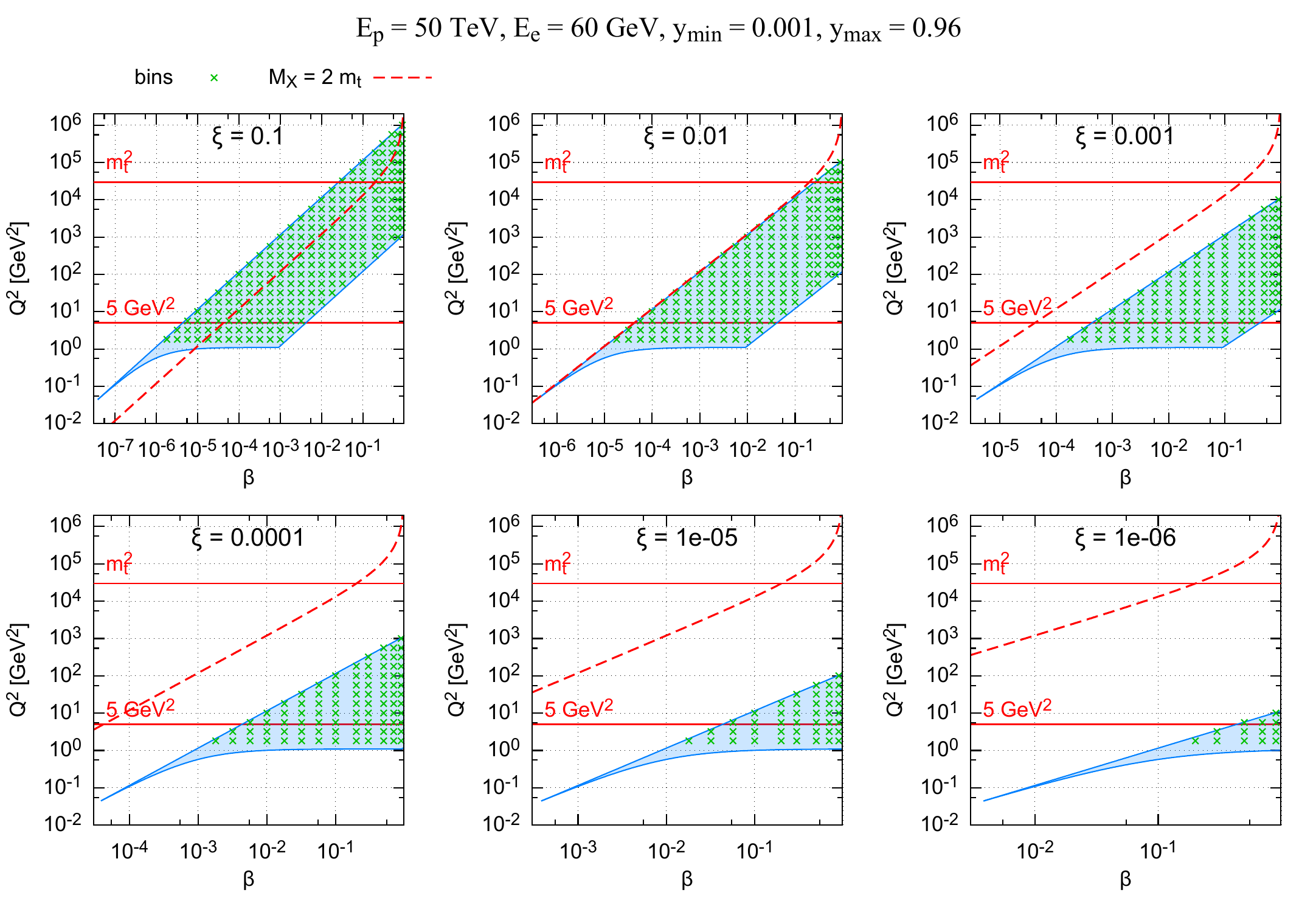}}
\caption{Kinematic phase space for inclusive diffraction in $(\beta,Q^2)$ for fixed values of $\xi$ for the FCC-eh design.
The horizontal lines indicate correspondingly, $Q_{\rm}^2=5 \; {\rm GeV}^2$, the lowest data value for the DGLAP fit performed in this study and $m_{\rm t}^2$ the 6-flavour threshold. The dashed line marks the kinematic limit for $t\bar t$ production.
}
\label{fig:phasespace_bQ_fcc}
\end{figure}

In \Fig{fig:phasespace_bQ_lhec} and \Fig{fig:phasespace_bQ_fcc} the phase space  in $(\beta,Q^2)$ is shown for fixed $\xi$ for the LHeC and FCC-eh, respectively. 
Both machines probe very small values of $\xi$, the LHeC 
reaching $10^{-4}$ with a wide range of $\beta$ and the FCC-eh 
extending $\xi$ down to $10^{-5}$.  Of course, the range in $\beta$ and $\xi$ 
is correlated since $x=\beta\xi$. Therefore for small values of $\xi$ only large values of $\beta$ are accessible while for large $\xi$ the range in $\beta$ extends to very small values. Above the solid, horizontal line labelled $m_t^2$ the top quark DPDF comes into play, and above the dashed line the $t\bar t$ production channel opens.


\subsection{Pseudodata for diffractive structure functions}
\label{sec:pseudo_data}

The reduced cross sections are extrapolated using Eqs.\Eq{eq:sred3} and \Eq{eq:FD3-fac} with the ZEUS-SJ DPDFs.
Following the scenario of the ZEUS fit \cite{Chekanov:2009aa} we work
within the VFNS scheme at NLO accuracy. The transition scales for 
DGLAP evolution are fixed by the heavy quark masses, $\mu^2 = m_h^2$
and the structure functions are calculated in the Thorne--Roberts GM-VFNS  \cite{Thorne:1997ga}.
The Reggeon PDFs are taken from the GRV pion set \cite{Gluck:1991ey},
the numerical parameters are taken from Tables 1 and 3 of Ref. \cite{Chekanov:2009aa}
and heavy quark masses are $m_c = 1.35\,\GeV, m_b = 4.3\,\GeV$,
and $\alpha_\mathrm{s}(M_Z^2) = 0.118$.

The model has a non-negligible Reggeon contribution which is 
hard to constrain from HERA data. It increases with increasing $\xi$ and gives 
a substantial contribution in the region $\xi>0.01$  for both the LHeC and the FCC-eh kinematics. Thus it is a source of a large uncertainty on the predictions in this region. 

The HERA kinematics give no access to the top quark region, and thus the model provides no reliable contributions from the top quarks. In the following simulations, 
the top quark contribution to the cross section is neglected, so that the extrapolated cross sections are likely underestimated for $Q^2>m_t^2$ and $M_X>2 m_t$
-- the significance of the top region is discussed in Sec.~\ref{sec:res}.

The pseudodata were generated using the extrapolation of the fit to HERA data, which provides the central values,
amended with a random Gaussian smearing 
with  standard deviation corresponding to the relative 
error $\delta$. An uncorrelated $5\%$ systematic error was assumed giving a total error
\be
\delta = \sqrt{\delta^2_{\rm sys}+\delta^2_{\rm stat}}\, .
\label{eq:uncertainty}
\ee
The statistical error was computed assuming a very modest  integrated luminosity of $2 \, {\rm fb}^{-1}$, see \cite{Bordry:2018gri,LHeClumi}. For the binning adopted in this study, the statistical uncertainties  have a very small effect on the uncertainties in the extracted DPDFs. Obviously, a much larger luminosity would allow a denser binning that would result in smaller DPDF uncertainties.

Note that our aim is not to provide a rigorous prediction of the reduced cross section or its full uncertainty.
Such a study would need to take account of theory uncertainties such as those stemming from the order in the 
perturbative expansion, the functional form of the initial conditions for DGLAP evolution and the values of the strong 
coupling constant and heavy quark masses. Our goal is simply to establish the extent to which data from new 
machines can reduce the existing experimental uncertainties in DPDFs relative to those from the fits to HERA data.

\begin{figure}
\centering{\includegraphics*[width=12cm,trim=0 5 0 20]{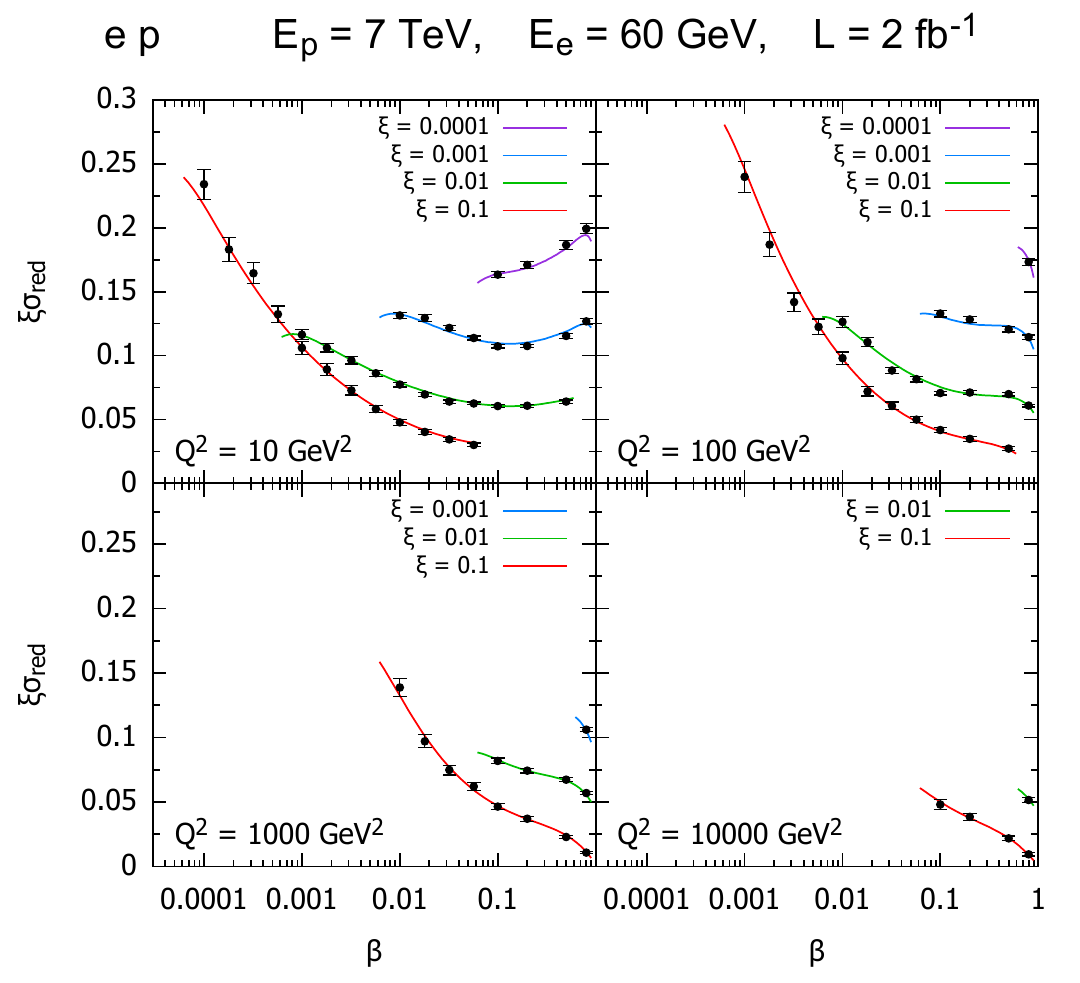}}
\caption{Selected subset of the  simulated data for the diffractive reduced cross section as a function of $\beta$ in bins of $\xi$ and $Q^2$ for $ep$ collisions at the LHeC.
The curves for $\xi = 0.01, 0.001, 0.0001$ are shifted up by 0.04, 0.08, 0.12, respectively.}
\label{fig:sigred_ep_lhec}
\end{figure}

\begin{figure}
\centering{\includegraphics*[width=12cm,trim=0 5 0 20]{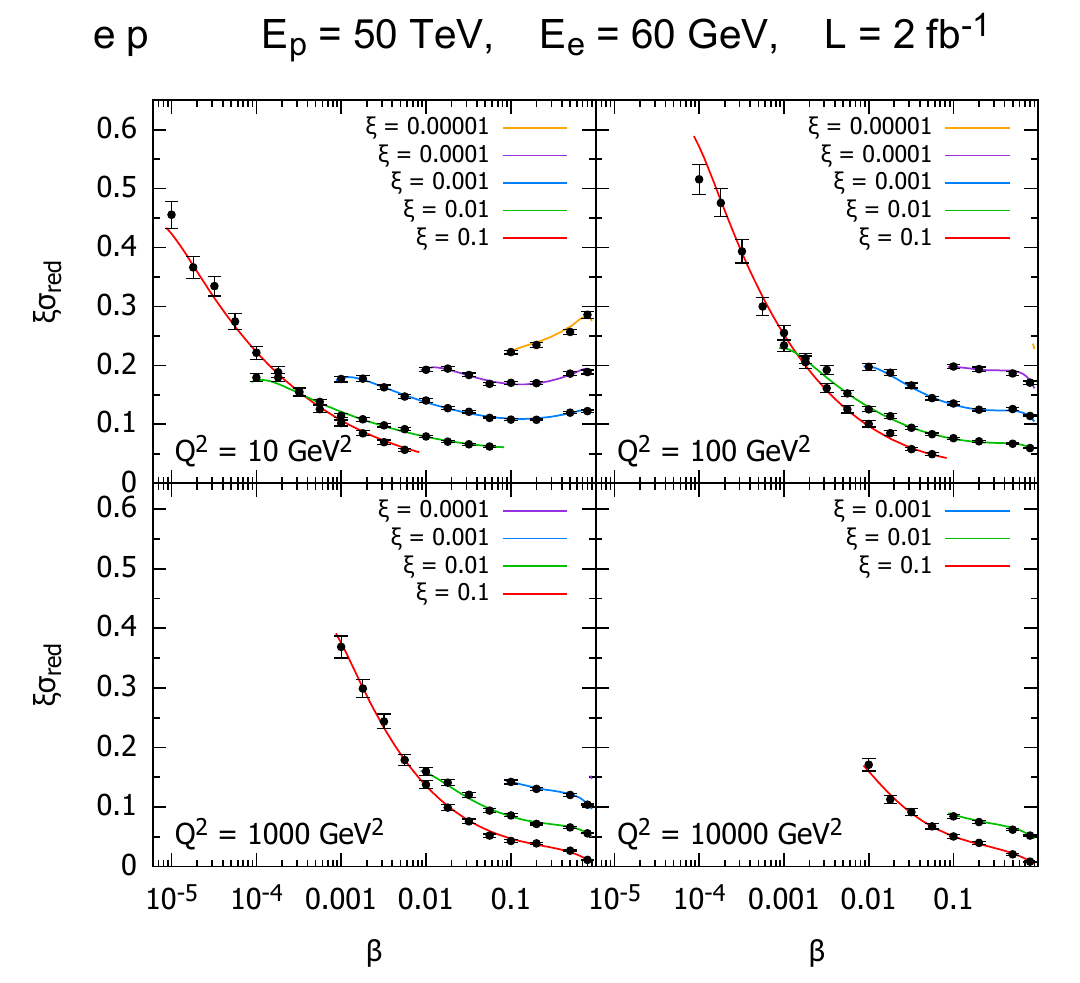}}
\caption{Selected subset of the  simulated data for the diffractive reduced cross section as a function of $\beta$ in bins of $\xi$ and $Q^2$ for $ep$ collisions at the FCC-eh.
The curves for $\xi = 0.01, 0.001, 0.0001, 0.00001$ are shifted up by 0.04, 0.08, 0.12, 0.16, respectively.}
\label{fig:sigred_ep_fcc}
\end{figure}

In \Fig{fig:sigred_ep_lhec} and \Fig{fig:sigred_ep_fcc} we show a 
subset of the simulated data for the diffractive reduced cross section 
$\xi\sigma_{\rm red}$ as a function of $\beta$ in selected bins of $\xi$ 
and $Q^2$ for the LHeC and FCC-eh cases, respectively. For 
the most part the errors are very small, and are dominated by the systematics. The breaking of Regge factorization evident at large $\xi$ comes from the large Reggeon contribution in that region, whose validity could be further investigated at the LHeC and FCC-eh. 


\section{Potential for constraining diffractive PDFs at the LHeC and  FCC-eh}
\label{sec:res}

\subsection{Fitting methodology and results}
\label{sec:methres}

With the aim of establishing the experimental precision with which DPDFs could be extracted when 
LHeC and FCC-eh data become available, we generate the central values of the pseudodata using the central set of the ZEUS-SJ fit that are distributed according to a Gaussian with experimental width, Eq. (\ref{eq:uncertainty}), that also provides the uncertainty in the pseudodata. 
We then include the pseudodata in a fit using the same functional form and, as expected,
obtain a $\chi^2/\mathrm{ndf} \sim 1$, which demonstrates the consistency of the approach.\footnote{As a 
cross-check of the method, we have performed a fit to the data simulated in the 
HERA kinematic region with HERA-like experimental errors $\sim 10\%$ 
and we recovered the ZEUS-SJ fit results and uncertainties with a very good accuracy. We have 
also modified the experimental uncertainties by a factor 2 and observed that the uncertainties in the extracted DPDFs are changed by the same factor.}
The fact that the $\chi^2$ is fully acceptable suggests
that using a more flexible form for each parton species, or adding more species by allowing parton decomposition, 
cannot improve the fit to the pseudodata in a meaningful way. 
Obviously, it may turn out when real data become available that the 
functional form used in our work is not able to produce a satisfactory fit and 
improvements of the parametrization would then be required, ideally with an assessment of the associated uncertainties. 
Understanding and quantifying such parametrization biases and uncertainties is a most important subject in its own right, see the comments at the end of Subsec. \ref{sec:uncertain}.,
that goes beyond the limited scope of this work.

To evaluate the precision with which the DPDFs can be determined,
several pseudodata sets,
corresponding to independent random error samples, were generated.
Each pseudodata set was fitted to the reduced cross-sections defined by
Eqs. \Eq{eq:sred3} and \Eq{eq:FD3-fac}
in the DPDF model of Sec.~\ref{sec:dpdf_param}.

The minimal value of $Q^2$ for the data considered in the fits 
was set to  $\QQmin = 5 \,\GeV^2$. The reason for this cut-off is  to show the feasibility of the fits including just the range in which standard twist-2 DGLAP evolution is expected to be trustable. At HERA, the $Q^2_{\rm min}$ values giving acceptable 
DGLAP (twist-2) fits were $8\,\GeV^2$ \cite{Aktas:2006hy}
and $5\,\GeV^2$ \cite{Chekanov:2008fh} for  H1 and ZEUS, respectively. It is expected that if there are
any higher twist effects, for example due to  parton saturation, they should become visible in the lower $Q^2$ region. 
DGLAP fits to the diffractive data are known to not describe the data very well in this region, which may point
to the  importance of the higher order or higher twist corrections. 

It is possible that a more flexible functional form would eventually be able to fit such data 
from the new machines without resorting to dynamics beyond twist-2 DGLAP
but, with the amount and precision of HERA data, no evidence for this was found.
Note that phenomenological studies which include higher twist corrections  indeed describe the HERA data in this region better than the pure DGLAP evolution \cite{Motyka:2012ty}.

The maximum value of $\xi$ was set 
by default to $\ximax = 0.1$, above 
which the cross-section starts to be dominated by the Reggeon exchange.
The effects of relaxing both limits \QQmin and \ximax are described below.
The region above the top threshold was not considered in the fits. 
This point however should be addressed in future studies; 
the top contribution has a negligible impact for the LHeC but some impact for the FCC-eh.

The binning adopted in this study corresponds roughly to 4 bins per order of magnitude in each of $\xi, \beta, Q^2$.
For $\QQmin = 5 \,\GeV^2$, $\ximax = 0.1$ and below the top threshold
this results in 1229 and 1735 pseudodata points for the LHeC and FCC-eh, respectively.
The top-quark region adds 17 points for the LHeC and 255 for FCC-eh.
Lowering $\QQmin$ down to $1.8\,\GeV^2$ we get 1589 and 2171 pseudodata points,
while increasing $\xi$ up to 0.32 adds ca. 180 points for both machines.

The potential for determination of the gluon DPDF was investigated by fitting the inclusive diffractive DIS pseudodata with two  models, S and C of Sec.~\ref{sec:dpdf_param}
with $\alpha_{\pom,\regg}(0)$ fixed, in order to focus on the shape of 
the Pomeron's PDFs.
At HERA, both S and C fits provide equally good 
descriptions of the data with $\chi^2/\mathrm{ndf} = 1.19$ and 1.18, respectively, 
despite different gluon DPDF shapes. 
The LHeC pseudodata are much more sensitive to gluons, resulting in
$\chi^2/\mathrm{ndf}$ values of 1.05 and 1.4 for the S and C fits, 
respectively. This motivates the use of the larger number of parameters in 
the fit-S model, which we employ in the further studies. It also shows clearly the potential of the LHeC and the FCC-eh to
better constrain the low-$x$ gluon and, therefore, unravel eventual departures from standard linear evolution.

\begin{figure}
\vspace*{3ex}
\centerline{\includegraphics*[width=.5\textwidth,trim=0 0 0 34]{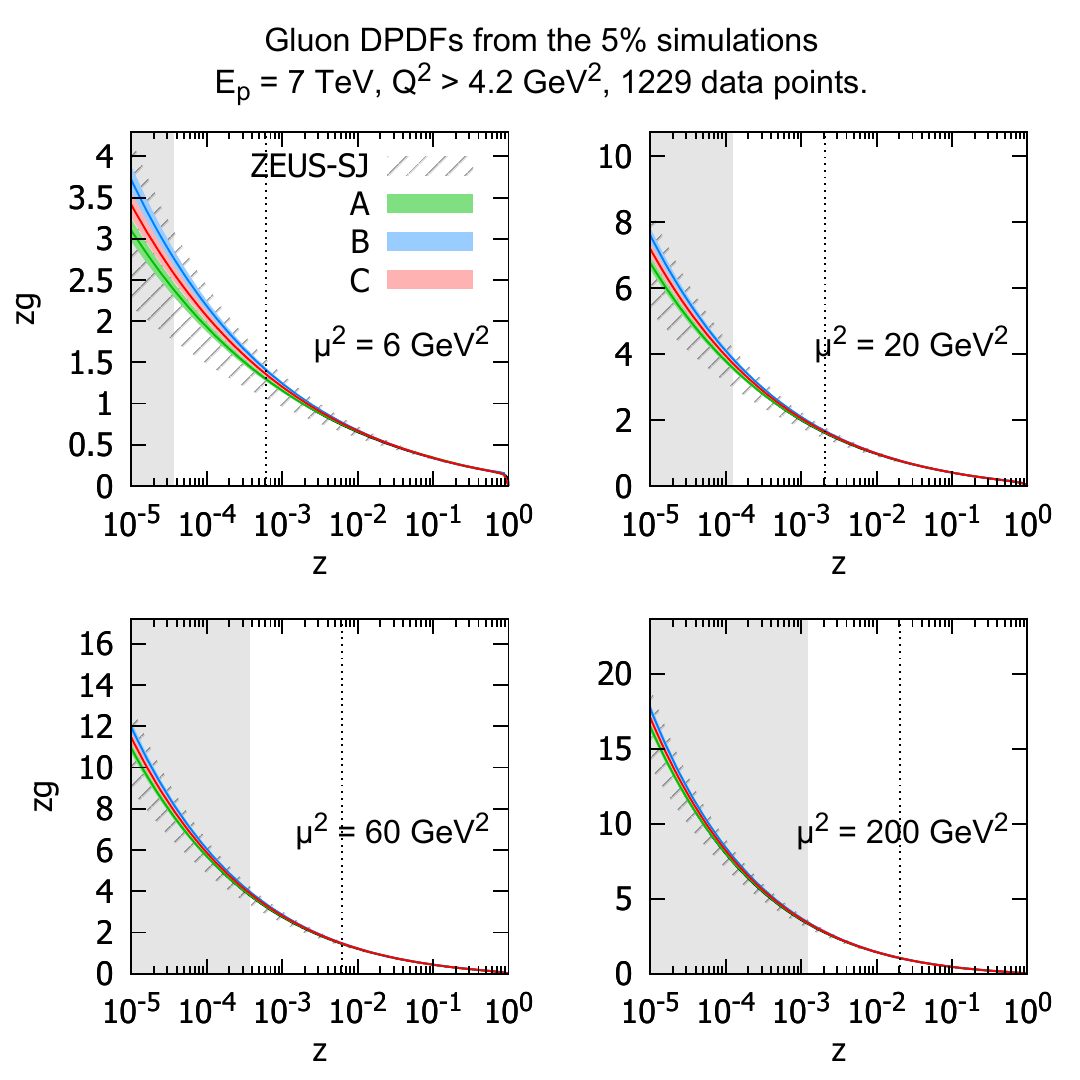}\!
\includegraphics*[width=.5\textwidth,trim=0 0 0 34]{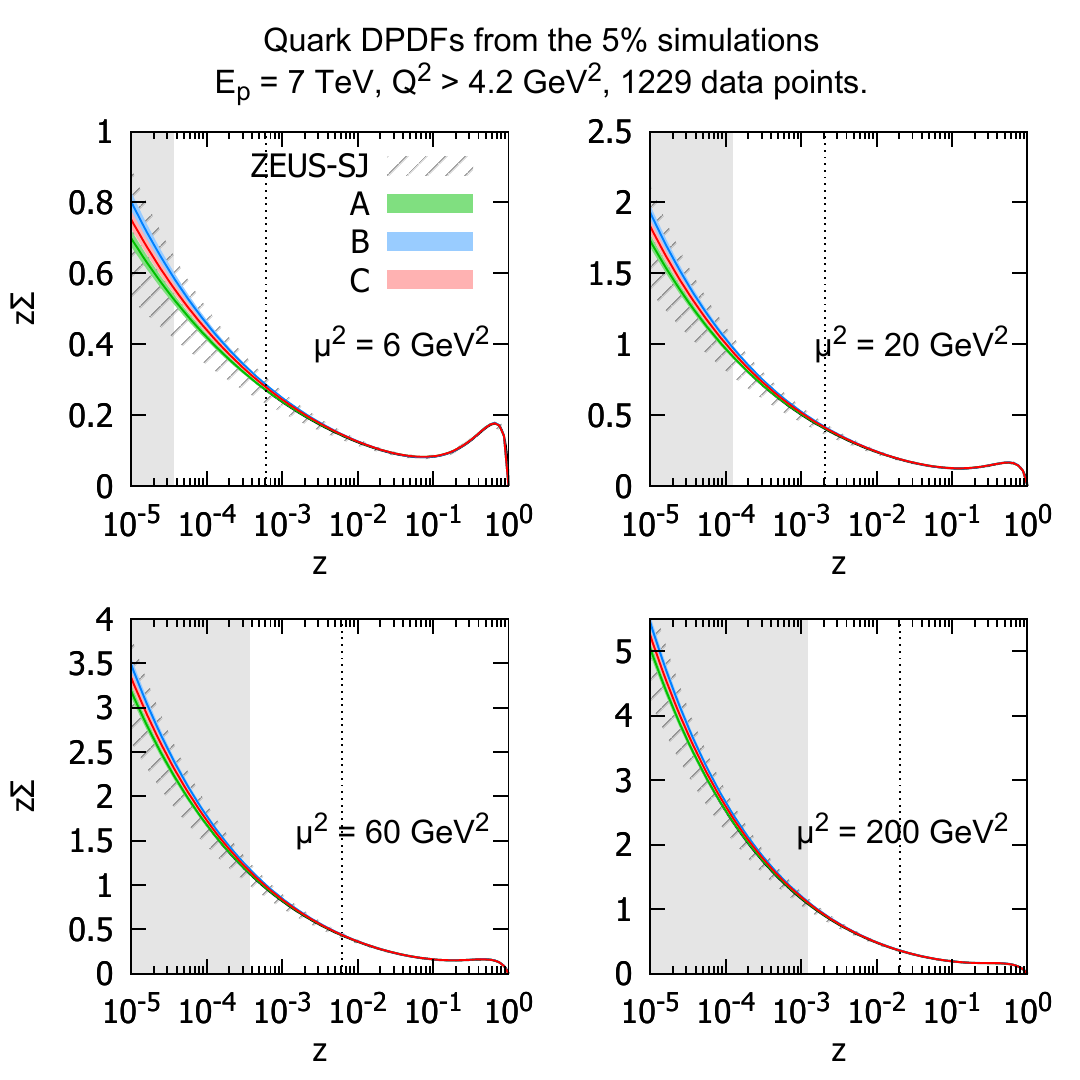}}
\caption{Diffractive PDFs for gluon and quark in the LHeC kinematics as a function of momentum fraction $z$ for fixed values of scale
$\mu^2$. Results of fits to three (A,B,C) pseudodata 
replicas are shown together with the experimental error bands.
For comparison, the extrapolated 
ZEUS-SJ fit is also shown (black) with error bands marked with the hatched pattern.
The vertical dotted lines indicate the HERA kinematic limit. The bands indicate only the experimental uncertainties, see the text.}
\label{fig:pdf_fits_lhec}
\end{figure}

\begin{figure}
\centerline{\includegraphics*[width=.5\textwidth,trim=0 0 0 34]{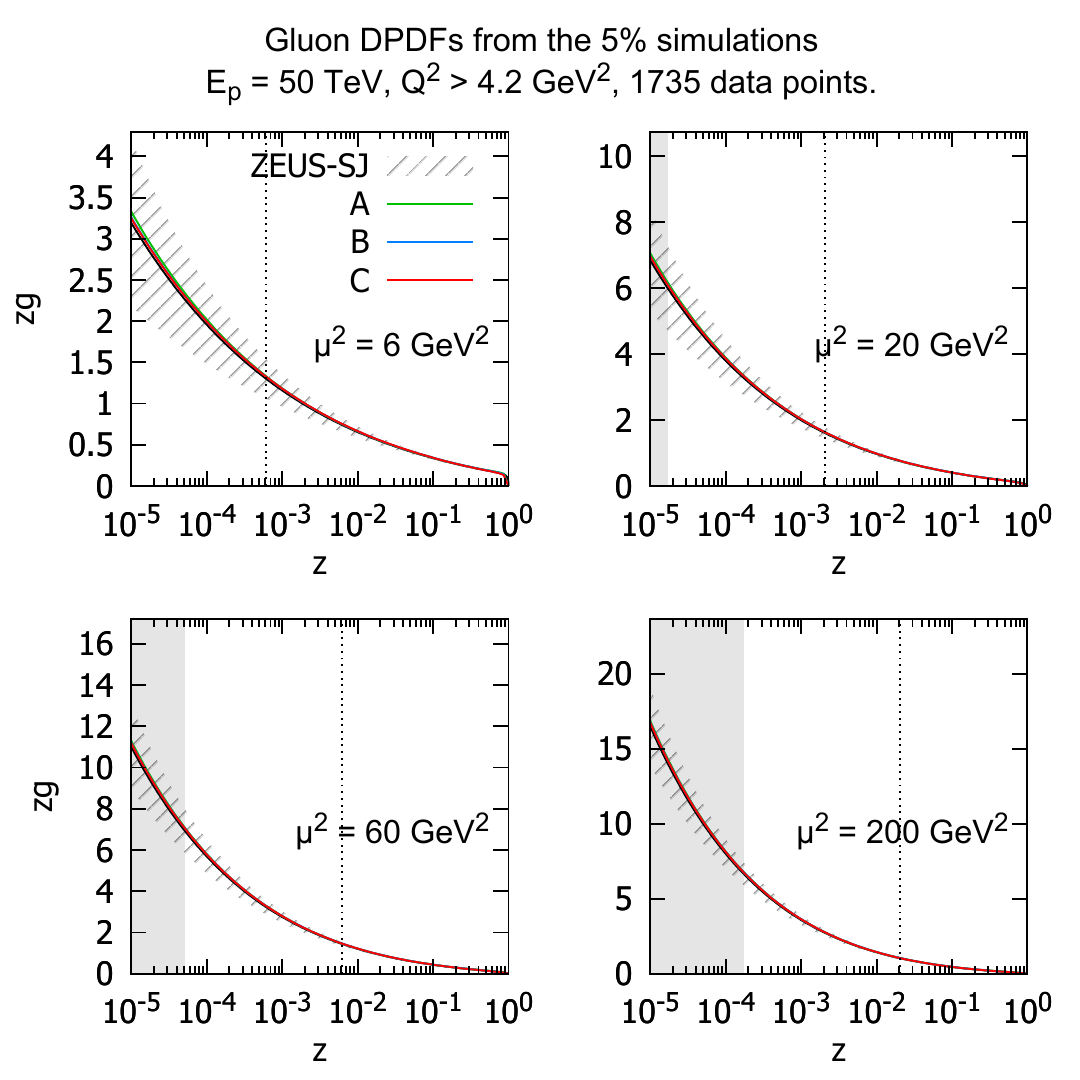}\!
\includegraphics*[width=.5\textwidth,trim=0 0 0 34]{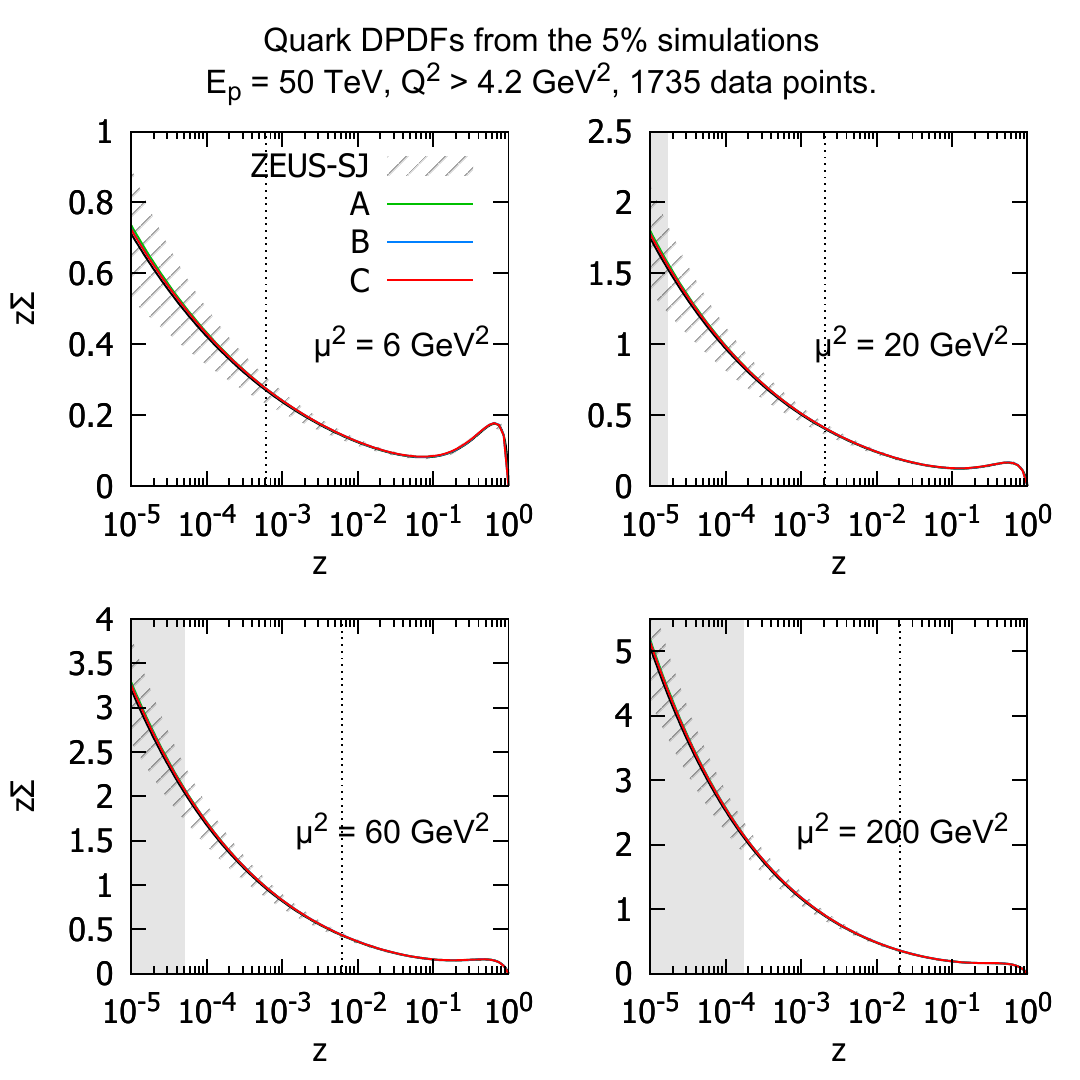}}
\caption{Identical to Fig. \ref{fig:pdf_fits_lhec}, but in the FCC-eh kinematics. The bands indicate only the experimental uncertainties, see the text.}
\label{fig:pdf_fits_fcc}
\end{figure}

\subsection{DPDFs uncertainties}
\label{sec:uncertain}

In \Fig{fig:pdf_fits_lhec} and \Fig{fig:pdf_fits_fcc} the diffractive gluon and quark 
distributions are shown for the LHeC and FCC-eh, respectively, as a  function of 
$z$ for fixed scales $\mu^2 = 6, 20, 60, 200\,\GeV^2$.
The bands labelled
$A,B,C$ denote fits to three statistically independent pseudodata replicas, obtained from the same central values and statistic and systematic uncertainties. Hereafter the bands
shown correspond to $\Delta\chi^2 = 2.7$ uncertainty (90\,\% CL).
Also the extrapolated ZEUS-SJ DPDFs are shown with error bands marked by 
the `/' hatched area.
Note that the depicted uncertainty bands come solely from experimental errors, neglecting 
theoretical sources, such as fixed input parameters and parametrization biases.
The extrapolation beyond the reach of LHeC/FCC-eh is marked in grey and the HERA kinematic limit is marked with the vertical dotted line. The stability of the results with respect to the replica used for the analysis is evident, so in the following only one will be employed.
The DPDFs determination accuracy improves with respect to 
HERA by a factor of 5--7 for the LHeC and 10--15 for the FCC-eh.

\begin{figure}
\centering{\includegraphics*[width=12cm,trim=0 0 0 48]{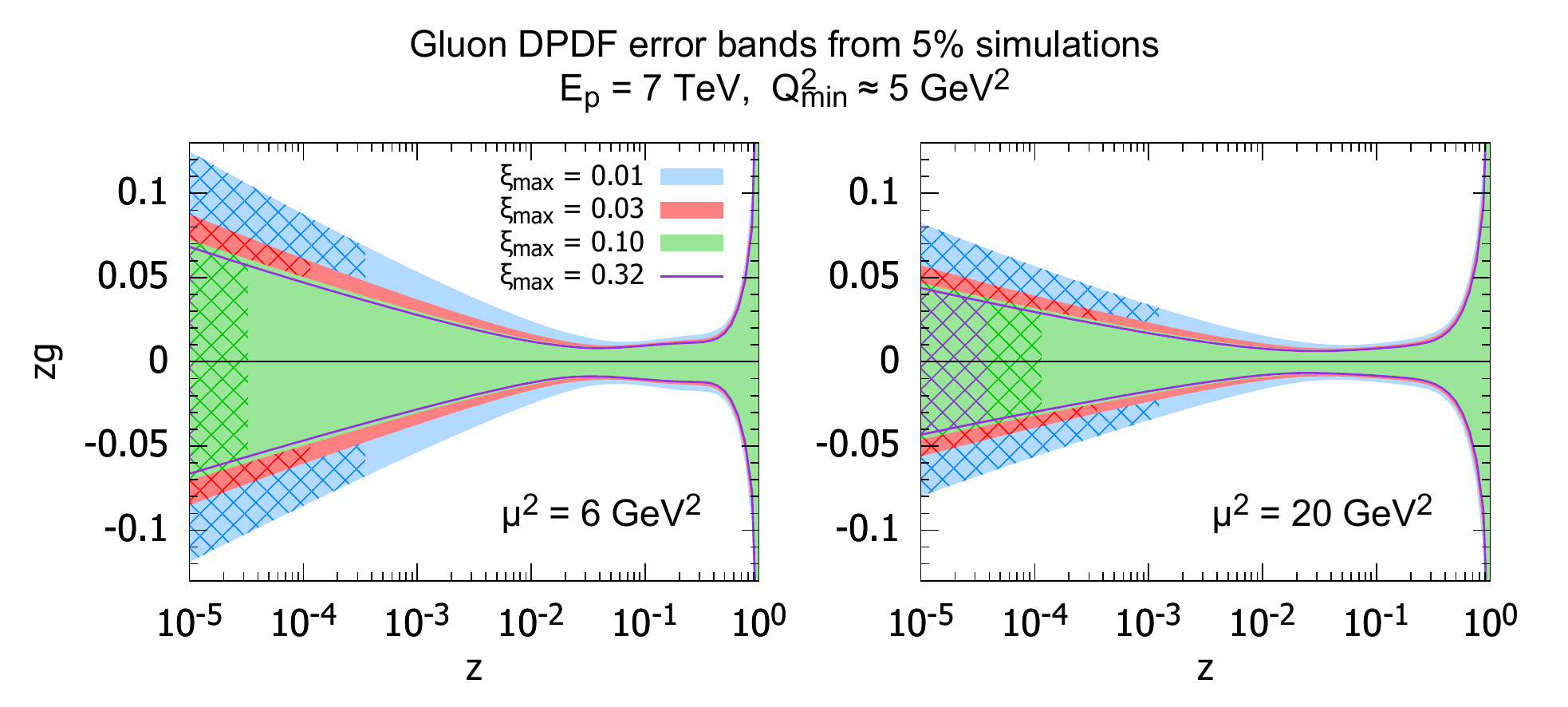}
\includegraphics*[width=12cm,trim=0 0 0 42]{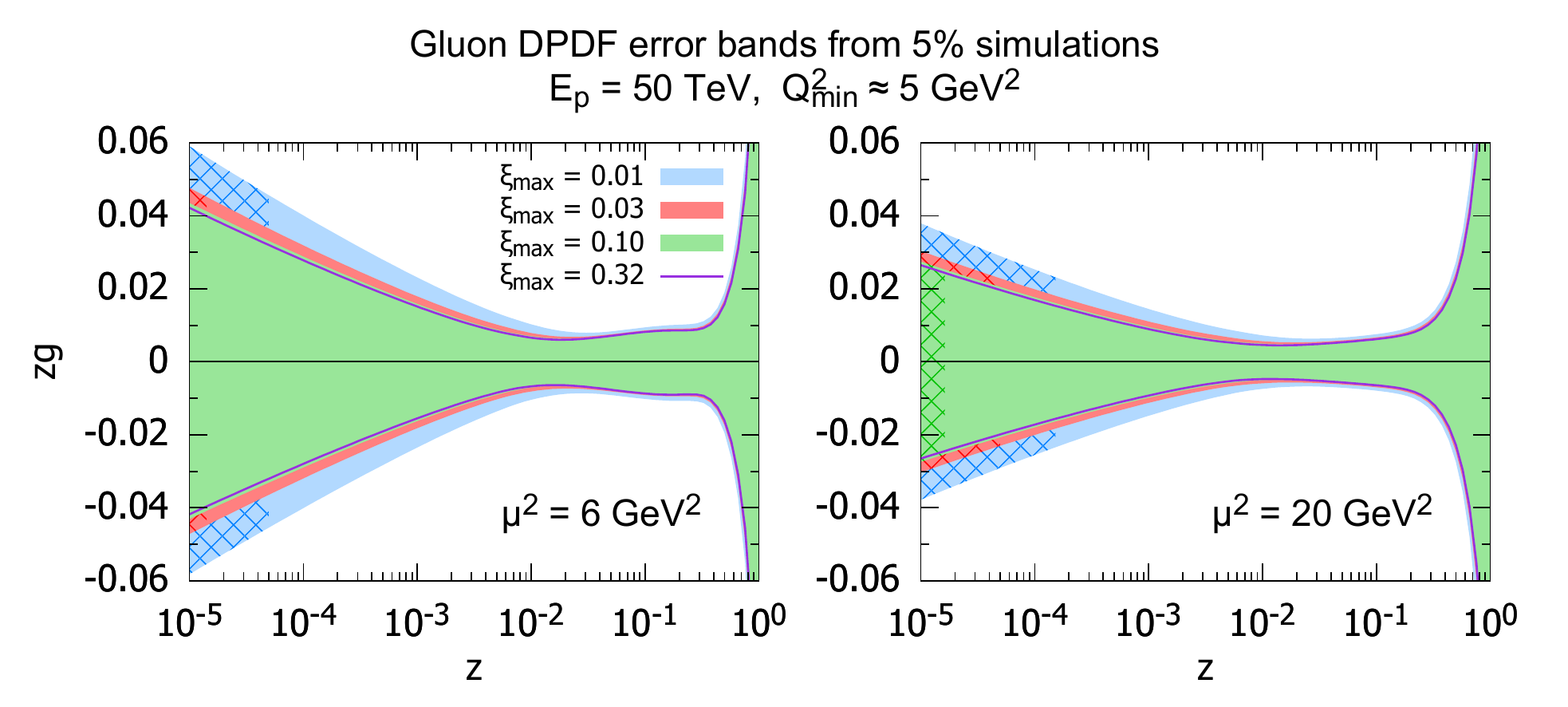}}
\caption{Relative uncertainties on the diffractive gluon PDFs for the LHeC kinematics (upper panel) and FCC-eh kinematics (lower panel). Two different choices of scales are considered $\mu^2=6$ and $\mu^2=20$ $\rm GeV^2$. The blue, red, green bands and magenta line correspond to different maximal values of $\xi = 0.01,0.03,0.1,0.32$, respectively.
The cross-hatched areas show kinematically excluded regions. The bands indicate only the experimental uncertainties, see the text.}
\label{fig:pdf_7_50_xi}
\end{figure}

\begin{figure}
\centering{\includegraphics*[width=10cm,trim=0 0 0 54]{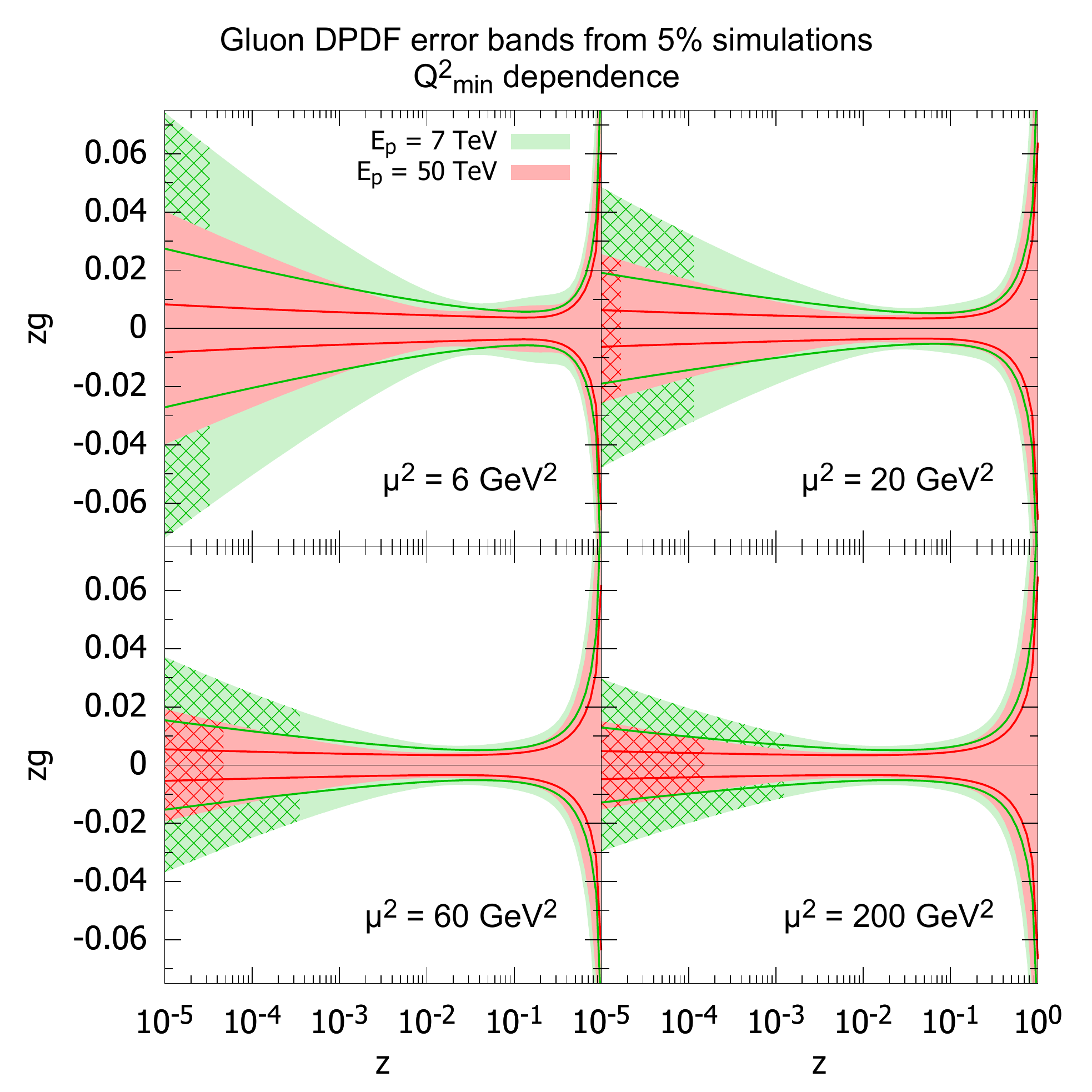}}
\caption{Relative uncertainties on the diffractive gluon PDF extraction for 
four distinct scales $\mu^2=6,20,60,200 \, \GeV^2$. 
The bands correspond to the choice of the high cut-off on the  
data included in the fit $Q^2_{\rm min}=5 \,\GeV^2$ and 
the lines correspond to the lower choice $Q^2_{\rm min}=1.8 \,\GeV^2$. 
The green colour corresponds to the LHeC scenario and red to the FCC-eh scenario.
The cross-hatched areas show kinematically excluded regions. The bands indicate only the experimental uncertainties, see the text.}
\label{fig:pdf_qHL}
\end{figure}

\begin{figure}
\centering{\includegraphics*[width=12cm,trim=0 0 0 44]{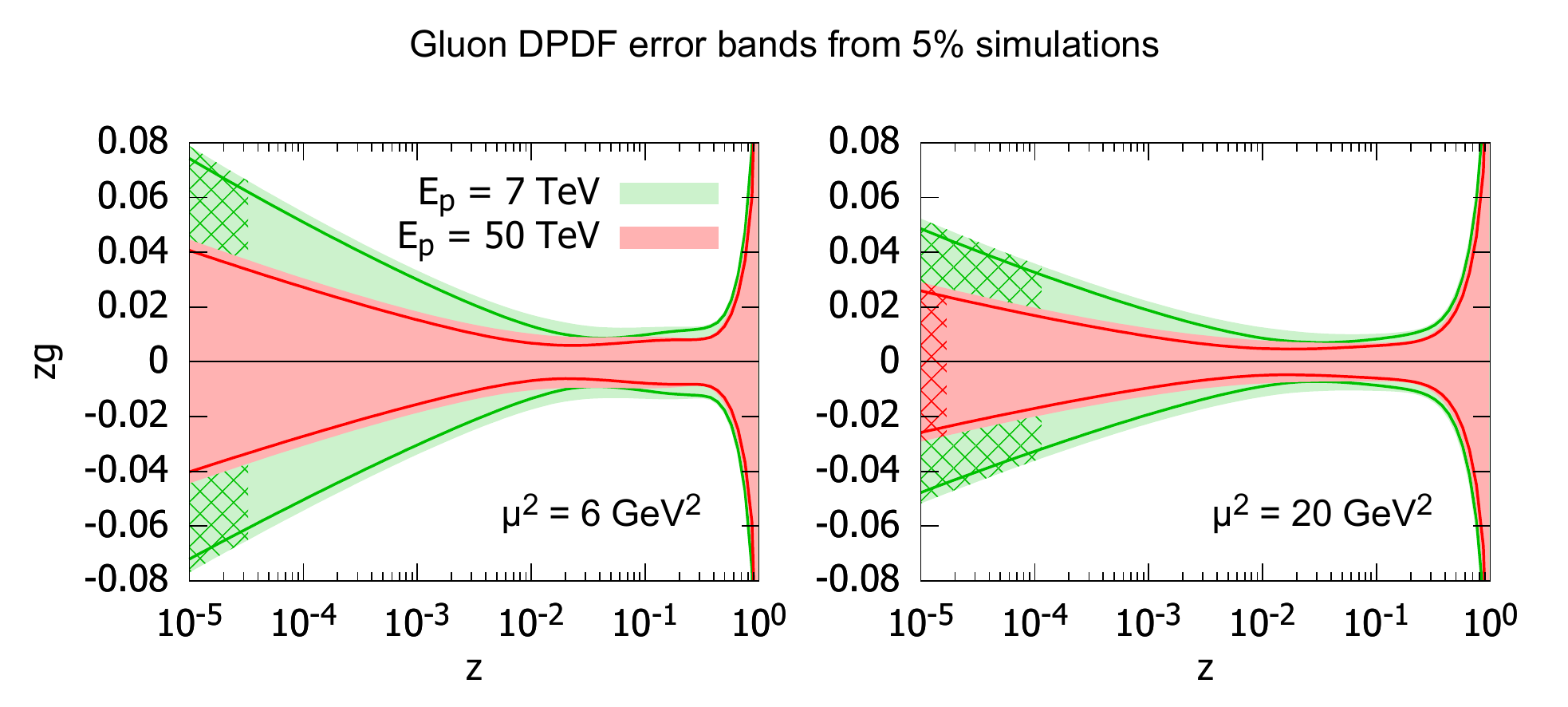}
\includegraphics*[width=12cm,trim=0 0 0 46]{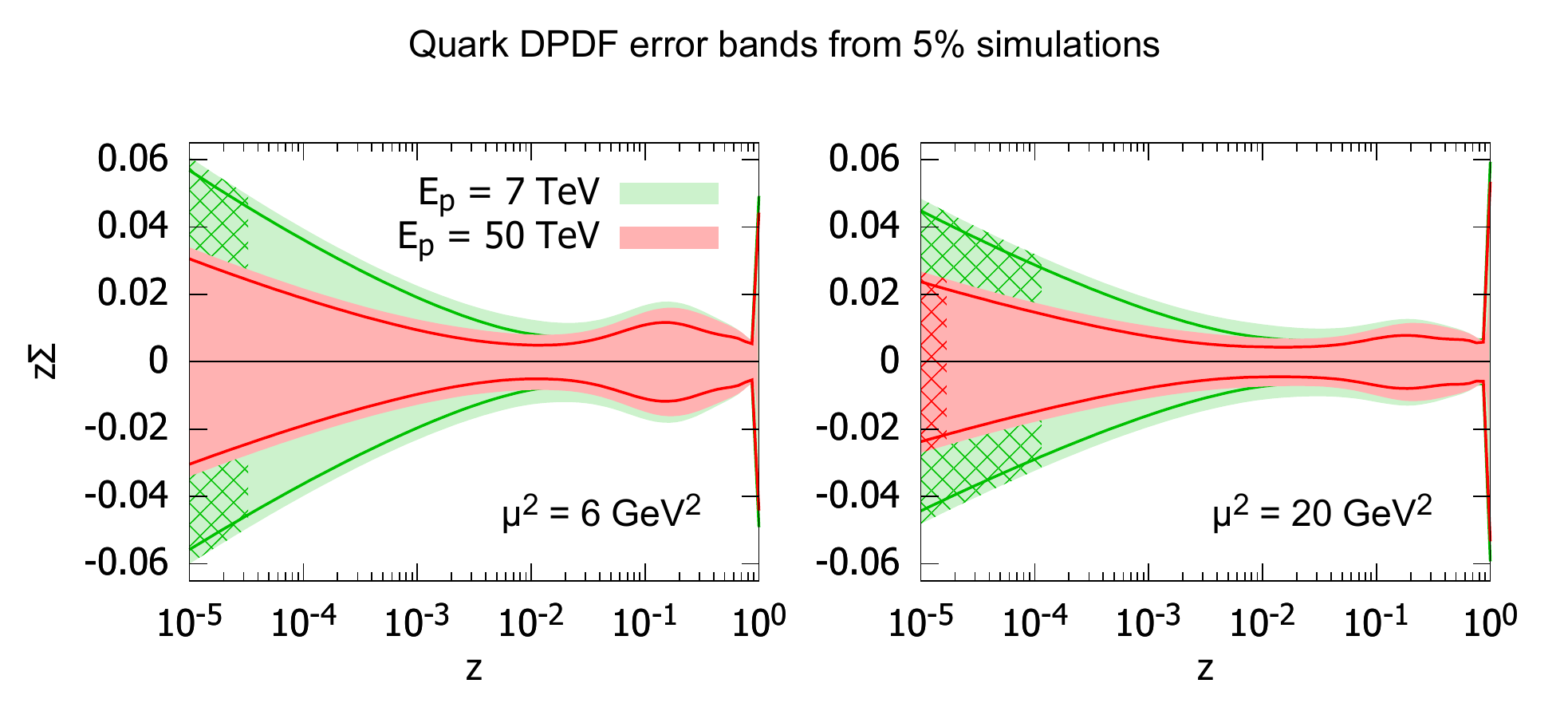}}
\caption{Relative uncertainties on the diffractive PDFs for different numbers 
of free fit parameters, 7 and 9. Two different 
choices of scales are considered $\mu^2=6$ and $20\,\GeV^2$.
The green and red bands correspond to the 9-parameter fits for the 
LHeC and FCC-eh scenarios, respectively. The continuous lines 
delimit the 7-parameter fit uncertainty.
The cross-hatched areas show kinematically excluded regions. The bands indicate only the experimental uncertainties, see the text.}
\label{fig:pdf-Npar}
\end{figure}

For a better illustration of the precision, 
in Figs.~\ref{fig:pdf_7_50_xi}, \ref{fig:pdf_qHL} and \ref{fig:pdf-Npar} 
the relative uncertainties  are shown for parton distributions at 
different scales.  In \Fig{fig:pdf_7_50_xi} the upper plots 
correspond to the LHeC and the lower ones to the FCC-eh scenarios, respectively. 
The different bands show the variation with the upper cut on the available 
$\xi$ range, from $0.01$ to $0.32$.  We observe only a modest 
improvement in the achievable accuracy of the extracted DPDFs with 
the change of $\xi$ by an order of magnitude from $0.01$ to $0.1$.  
An almost negligible effect is observed when further extending 
the $\xi$ range up to $0.32$.  This is encouraging, 
since the measurement for the very large values of $\xi$ is challenging. It reflects
the dominance of the secondary Reggeon in this region.

In \Fig{fig:pdf_qHL}  we show the variation of the relative precision 
with the change of the minimal value of $Q^2$ from $1.8 \ {\rm GeV^2}$ 
(curves) to $5\,\GeV^2$ (bands).   The LHeC scenario is indicated in green and 
FCC-eh in red. 
There is a quite substantial effect on the achieved 
precision depending on the minimal value of  $Q^2$. This is 
not only related to the fact that the number of pseudodata points is larger 
by about 300 in each case, but is primarily due 
to the fact that acceptance across the full range of $z$ in
this region is crucial for constraining  the initial condition for the DGLAP evolution. The more data points are in the region closer to the starting distribution the better it is constrained, particularly at low and medium values of $Q^2$ and $z$.   
\Fig{fig:pdf_qHL} also demonstrates that both machines will be very sensitive to this region and  therefore potentially able to constrain higher twists and/or saturation effects.

In \Fig{fig:pdf-Npar} we show the effect on 
the relative uncertainties for quarks and gluons of making 
$\alpha_{\pom,\regg}(0)$  free fit parameters. The 
increased number of fitting parameters from 7 to 9 has a very small 
effect on the DPDF uncertainties. In addition, we note that for low $x$ values the quark and gluon uncertainties are similar, with quark uncertainties being  smaller by about $20\%$. 
There is, however, a marked difference in 
the uncertainties for quarks and gluons at large values of $z$. 

As a concluding comment in this Sec., let us briefly discuss the influence of the functional form of the initial conditions for DGLAP evolution on the uncertainties obtained from fitting pseudodata. For this purpose, the following exercise has been performed. We have  included four additional parameters in \eqref{eq:initcond} through multiplying it by $1+D_i z +E_i \sqrt{z}$, for $i$ equal to a gluon or a light quark. We have checked that such form, with given values of $D_i,E_i$, is not in large disagreement with the HERA diffractive cross sections. This form and choice of parameters has  been used to generate pseudodata in the HERA and LHeC kinematics. These pseudodata have first been fitted using our standard ZEUS-SJ form, $D_i=E_i=0$, and then including $D_i,E_i$ one by one as additional fitting parameters. We generically observe that, in both kinematic domains, the improvement in the $\chi^2$/ndf when increasing the number of parameters is marginal - from one per cent to less than one per thousand. And that the size of the uncertainty bands is not larger than in the ZEUS-SJ based analysis. Therefore, in the kinematic ranges that we are exploring and with this functional form, we have been unable to quantify a meaningful parametrisation uncertainty. Obviously, these results and conclusions are linked to a given functional form and approach to estimate the uncertainties. A different functional form or the use of e.g. the NNPDF approach \cite{DelDebbio:2004xtd} instead of the Hessian method \cite{Pumplin:2001ct}, may be essayed. But the answers obtained using these different forms and approaches would be, as in our case, linked to the specific choices and methodologies. We take these facts as indicative that a proper treatment of parametrisation uncertainties can only be addressed on the basis of real data and supportive of our strategy here of focusing solely on the experimental uncertainties.


\section{Diffractive deep inelastic scattering off nuclei}
\label{sec:nuclei}
Electron-nucleus ($e$A) collisions are also possible at the LHeC 
and the FCC-eh with large integrated luminosities, ${\cal L}_{NN}\sim {\cal O}(1)$ fb$^{-1}$, 
see \cite{Dainton:2006wd,AbelleiraFernandez:2012cc,Klein:2018rhq,Bordry:2018gri,LHeClumi}. 
Similar considerations apply to diffraction in $e$A as to $ep$ collisions. 
The main difference is the larger contribution from 
incoherent diffraction\footnote{$A^*$ denotes a final state in which the nucleus has dissociated to a system of at least two hadrons, but the rapidity gap signature that 
defines the diffractive event is still present.} $e+A \to e+X+A^*$ than from coherent diffraction $e+A \to e+X+A$, the former dominating for $|t|$ larger than a few hundredths of a 
GeV$^2$.  In the following we focus on coherent diffraction,
which could be distinguished from the incoherent case 
using forward detectors \cite{AbelleiraFernandez:2012cc}.

Assuming the same framework (collinear factorization for hard 
diffraction, Eq. \eqref{eq:collfac}, and Regge factorization,
Eq. \eqref{eq:param_2comp}) described for $ep$ in 
Sections \ref{sec:sec2} and \ref{sec:dpdf_param} to 
hold for $e$A, nuclear diffractive PDFs (nDPDFs) can be 
extracted from the diffractive reduced cross sections, Eqs. \eqref{eq:sigmared4} 
and  \eqref{eq:sigmared3}. It should be noted that such nDPDFs have never been measured. With the same electron energy $E_e=60$ GeV and nuclear beams with $E_N=2.76$ and 19.7 TeV/nucleon for the LHeC and the FCC-eh, 
respectively, the kinematic 
coverage is very similar to that shown in Fig.~\ref{fig:phasespace_xQ}.

Due to the lack of previous measurements, there are no parametrizations for nDPDFs but models exist  for the nuclear effects on 
parton densities defined through the nuclear modification factor
\be
R_k^A(\beta,\xi,Q^2) = \frac{f_{k/A}^{D(3)}(\beta,\xi,Q^2)}{A\,f_{k/p}^{D(3)}(\beta,\xi,Q^2)}\;,
\label{eq:ndmf}
\ee
with 
diffractive parton densities in nucleus $A$, $f_{k/A}^{D(3)}(\beta,\xi,Q^2)$.
We use the model proposed in \cite{Frankfurt:2011cs}, where 
parametrizations for nuclear modification factors 
are provided at the scale $Q^2=4$ GeV$^2$ (extended in $\beta$ and $\xi$ to cover the LHeC and FCC-eh kinematic regions\footnote{We thank Vadim Guzey for providing them.}). Then DGLAP evolution is employed to
evolve the ZEUS-SJ proton diffractive PDFs multiplied by $R_k^A$ from \cite{Frankfurt:2011cs}
to obtain the nuclear diffractive PDFs, at any $Q^2$.
The structure functions and reduced cross sections are then
calculated in the same way as in the proton case, and these results are used to obtain the modification factors, analogous to Eq. \Eq{eq:ndmf}, 
for these quantities.
We have also repeated the calculation in the Zero-Mass VFNS in order to 
check that the resulting modification factors do not depend on the applied scheme.

\begin{figure}
\centering{\includegraphics*[width=12cm,trim=0 0 0 11]{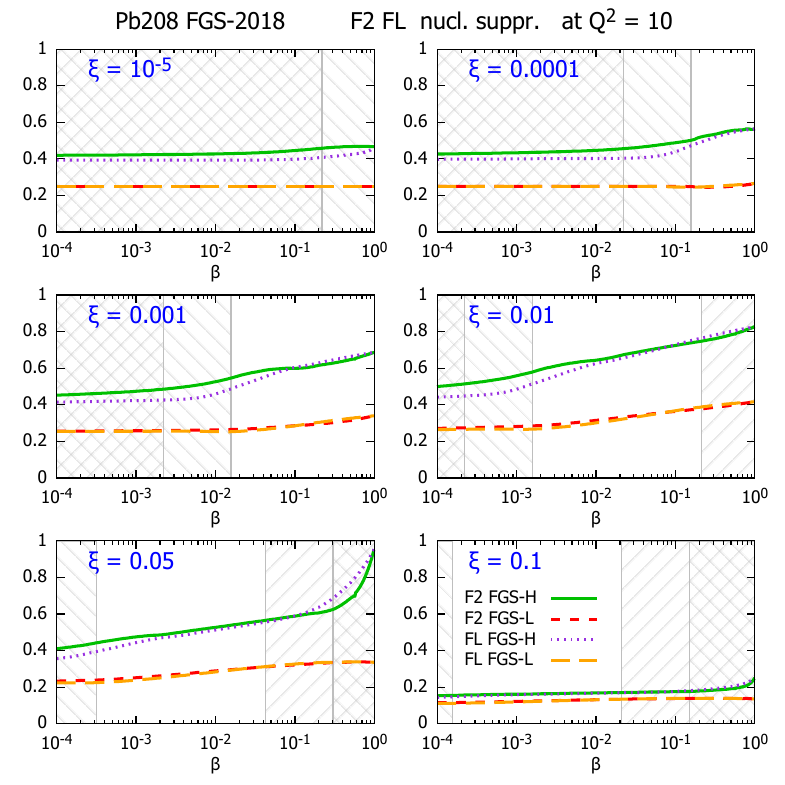}}
\caption{Nuclear modification factor, Eq. \eqref{eq:ndmf}, for $F_{2}^{D(3)}$ 
and $F_{L}^{D(3)}$ in $^{208}$Pb versus $\beta$, at $Q^2=10$ GeV$^2$ and 
for different $\xi$, for the models H and L in \cite{Frankfurt:2011cs}. 
The `\textbackslash' and `/' hatched areas show 
kinematically excluded regions for $E = 2.76$ and 19.7 TeV/nucleon, respectively.}
\label{fig:ratio_Pb50e60}
\end{figure}

The model in \cite{Frankfurt:2011cs} employs 
Gribov inelastic shadowing \cite{Gribov:1968jf} which relates diffraction in 
$ep$ to nuclear shadowing for total and diffractive 
$eA$ cross sections. It assumes that the nuclear wave function squared can be approximated by the product of one-nucleon densities, neglects the $t$-dependence of the diffractive $\gamma^*$-nucleon amplitude compared to the nuclear form factor, introduces a real part in the amplitudes \cite{Gribov:1968uy}, and considers the colour fluctuation formalism for the inelastic intermediate nucleon states \cite{Frankfurt:1994hf}. 
There are two variants of the model, named H and L, corresponding to different strengths of the colour fluctuations, giving rise to larger and smaller probabilities for diffraction in nuclei with respect to that in proton, respectively.
To illustrate the results of this model, in Fig.~\ref{fig:ratio_Pb50e60} 
we show the nuclear modification factor, Eq. \eqref{eq:ndmf}, 
for $F_{2}^{D(3)}$ and $F_{L}^{D(3)}$ in $^{208}$Pb.

\begin{figure}
\centerline{%
	\includegraphics*[width=0.5\textwidth,trim=0 0 0 24]{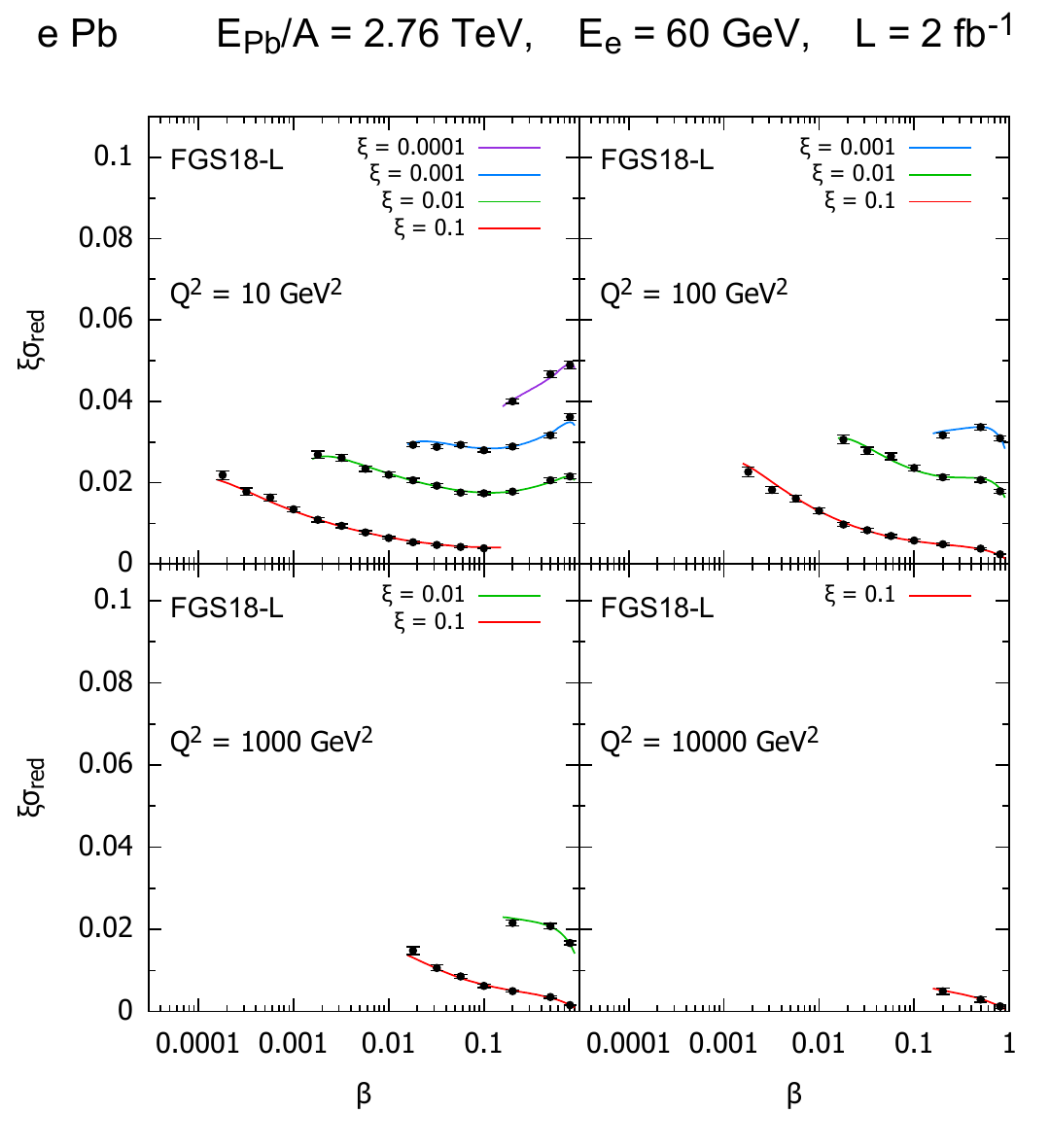}%
	\hspace*{-5pt}%
	\includegraphics*[width=0.5\textwidth,trim=0 0 0 24]{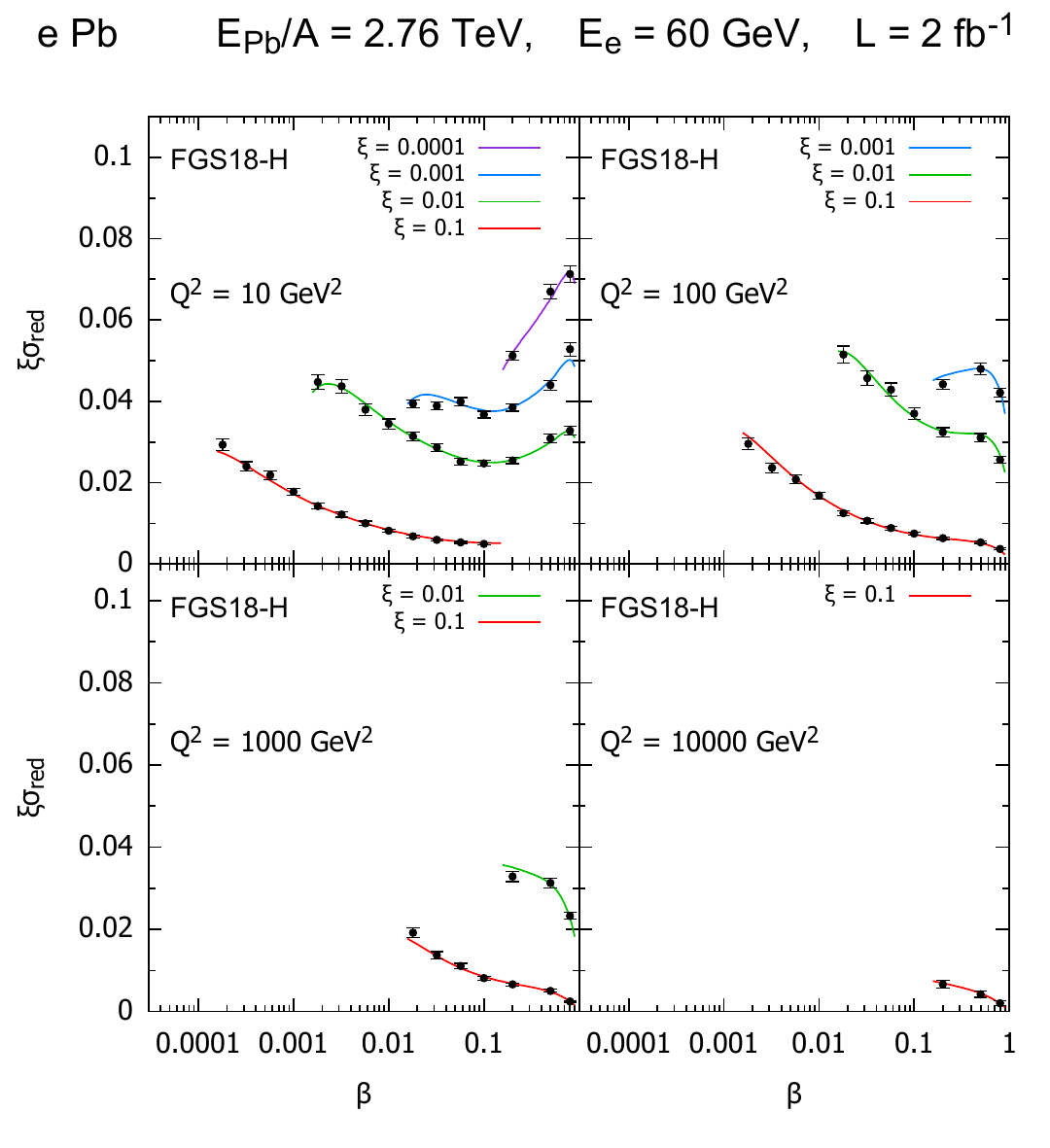}%
}
\caption{Simulated data for the diffractive reduced cross section as a function of $\beta$ in bins of $\xi$ and $Q^2$ for $e\,^{208}$Pb collisions at the LHeC, in the models in \cite{Frankfurt:2011cs}.
The curves for $\xi = 0.01, 0.001, 0.0001$ are shifted up by 0.01, 0.02, 0.03, respectively.}
\label{fig:sigred_Pb7e60}
\end{figure}

\begin{figure}
\centerline{%
	\includegraphics*[width=0.5\textwidth,trim=0 0 0 24]{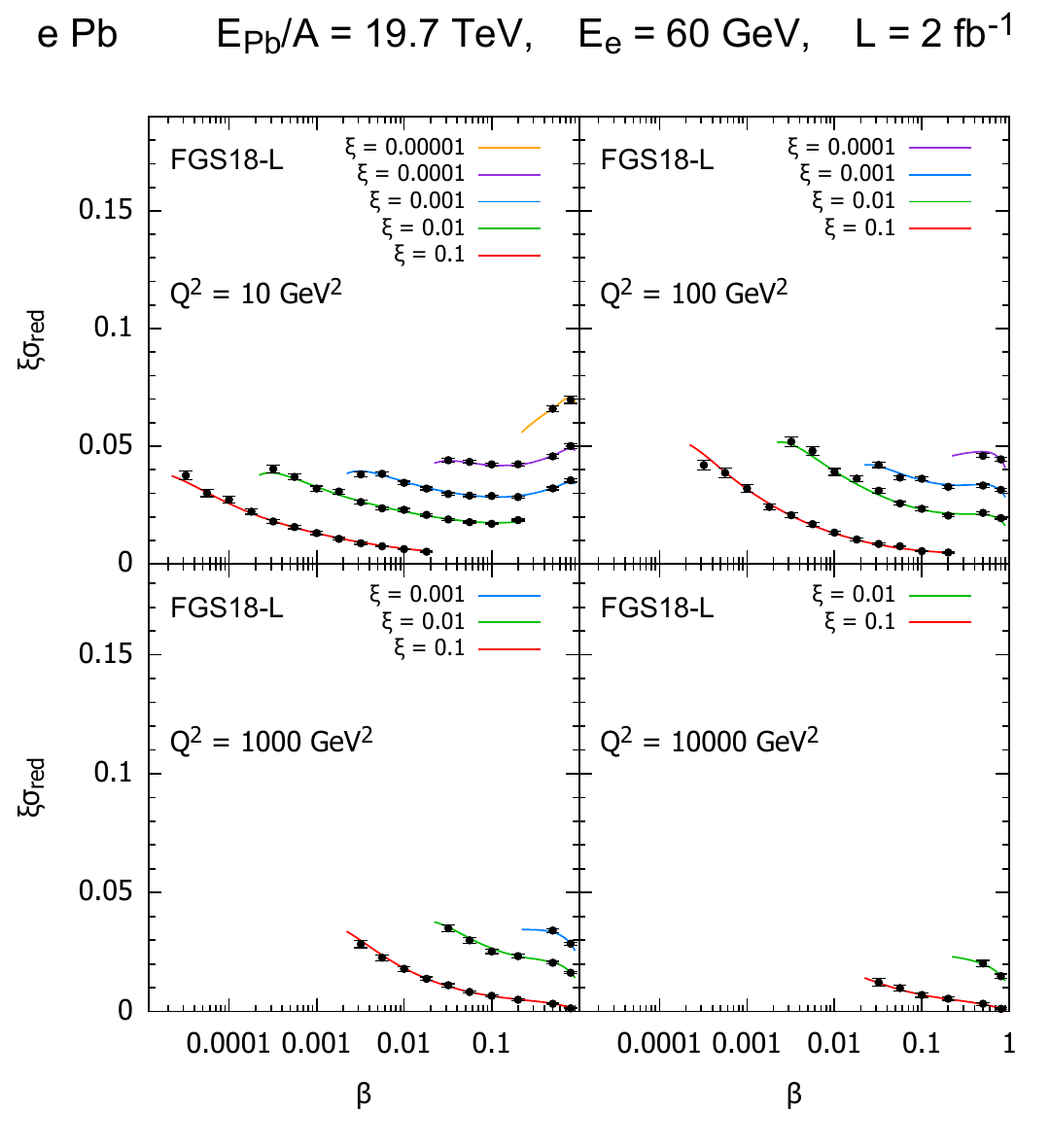}%
	\hspace*{-5pt}%
	\includegraphics*[width=0.5\textwidth,trim=0 0 0 24]{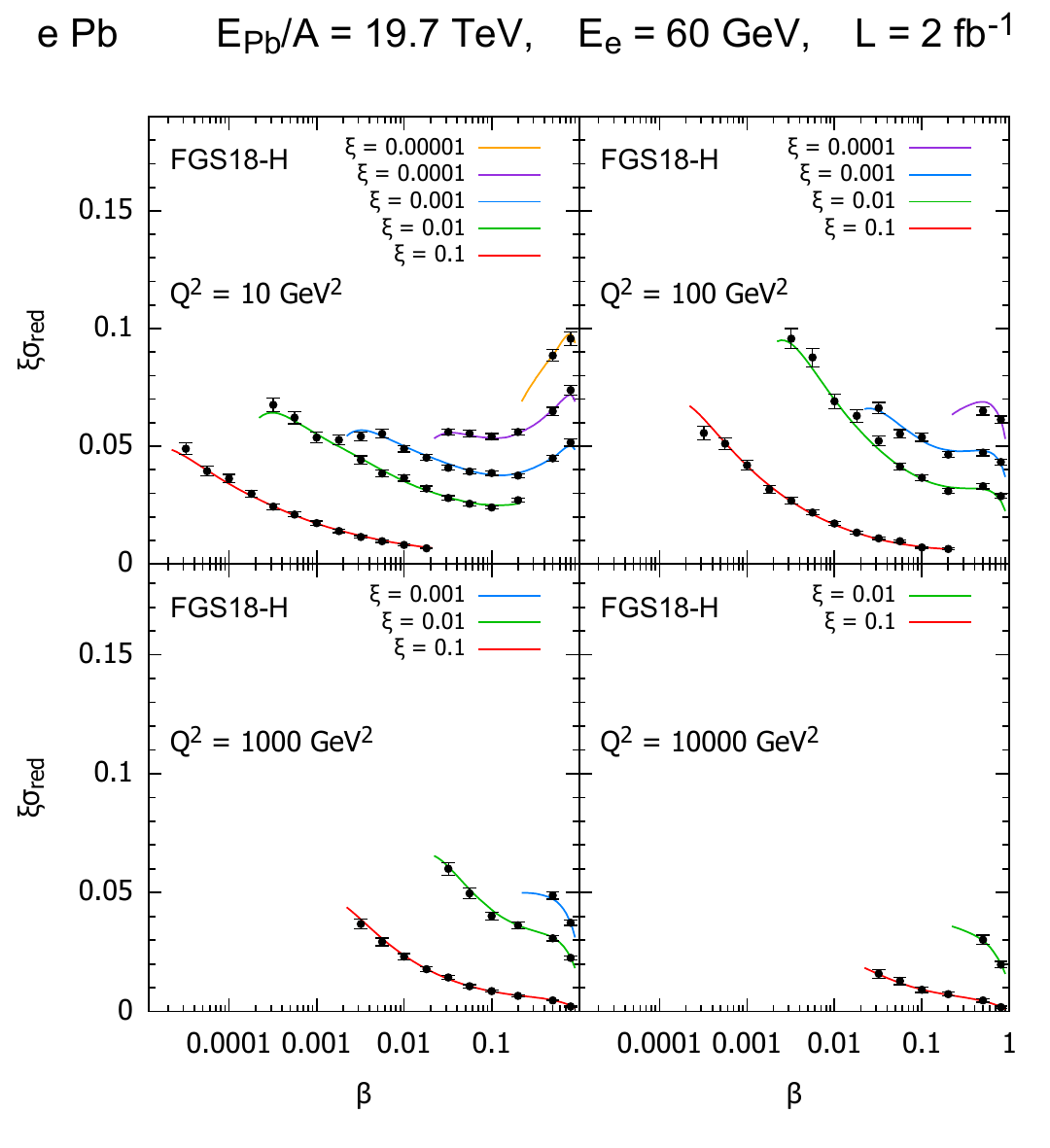}%
}
\caption{Simulated data for the diffractive reduced cross section as a function of $\beta$ in bins of $\xi$ and $Q^2$ for $e\,^{208}$Pb collisions at the FCC-eh, in the models in \cite{Frankfurt:2011cs}.
The curves for $\xi = 0.01, 0.001, 0.0001, 0.00001$ are shifted up by 0.01, 0.02, 0.03, 0.04, respectively.}
\label{fig:sigred_Pb50e60}
\end{figure}

\begin{figure}
\centerline{%
	\includegraphics*[width=0.5\textwidth,trim=0 0 0 30]{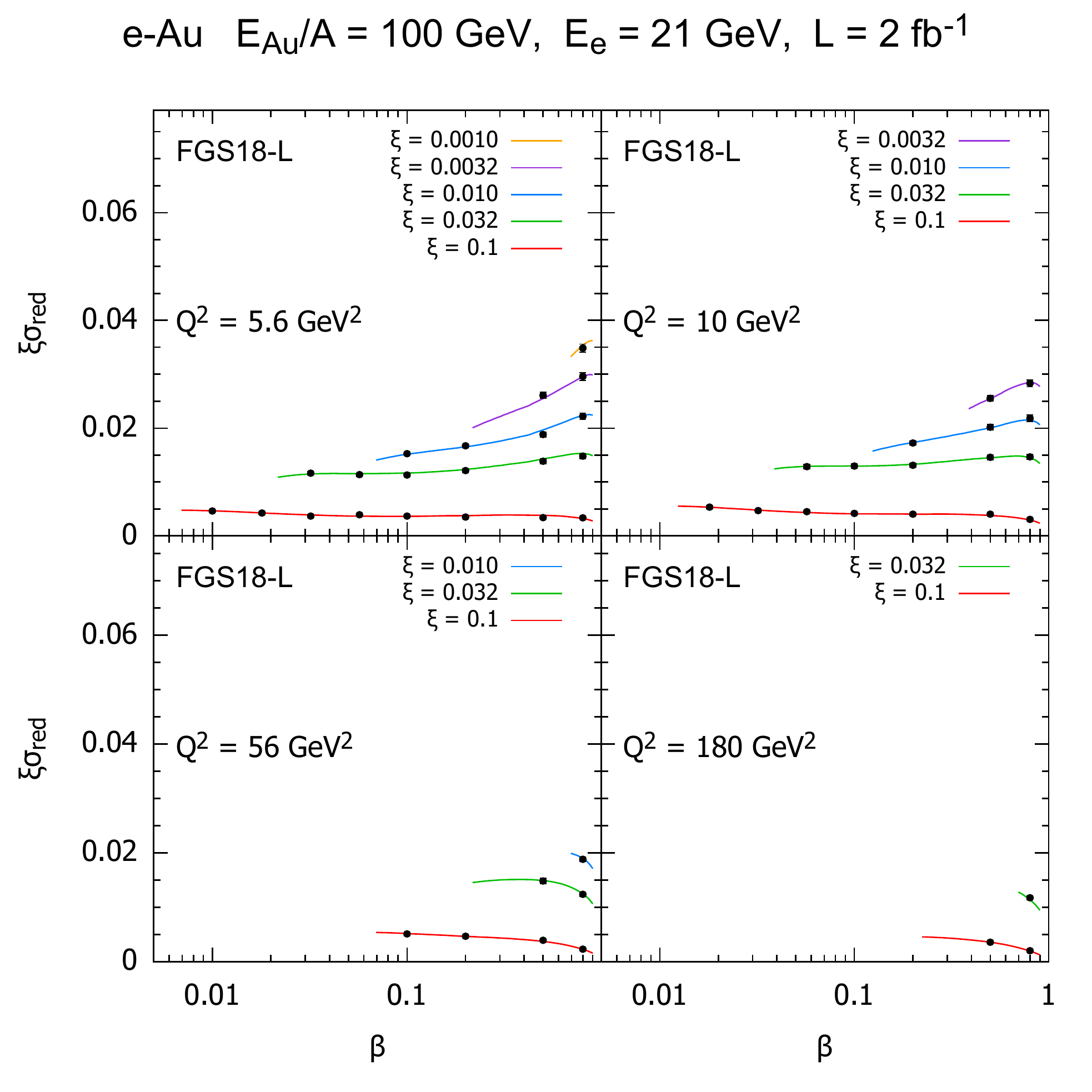}%
	\hspace*{-5pt}%
	\includegraphics*[width=0.5\textwidth,trim=0 0 0 30]{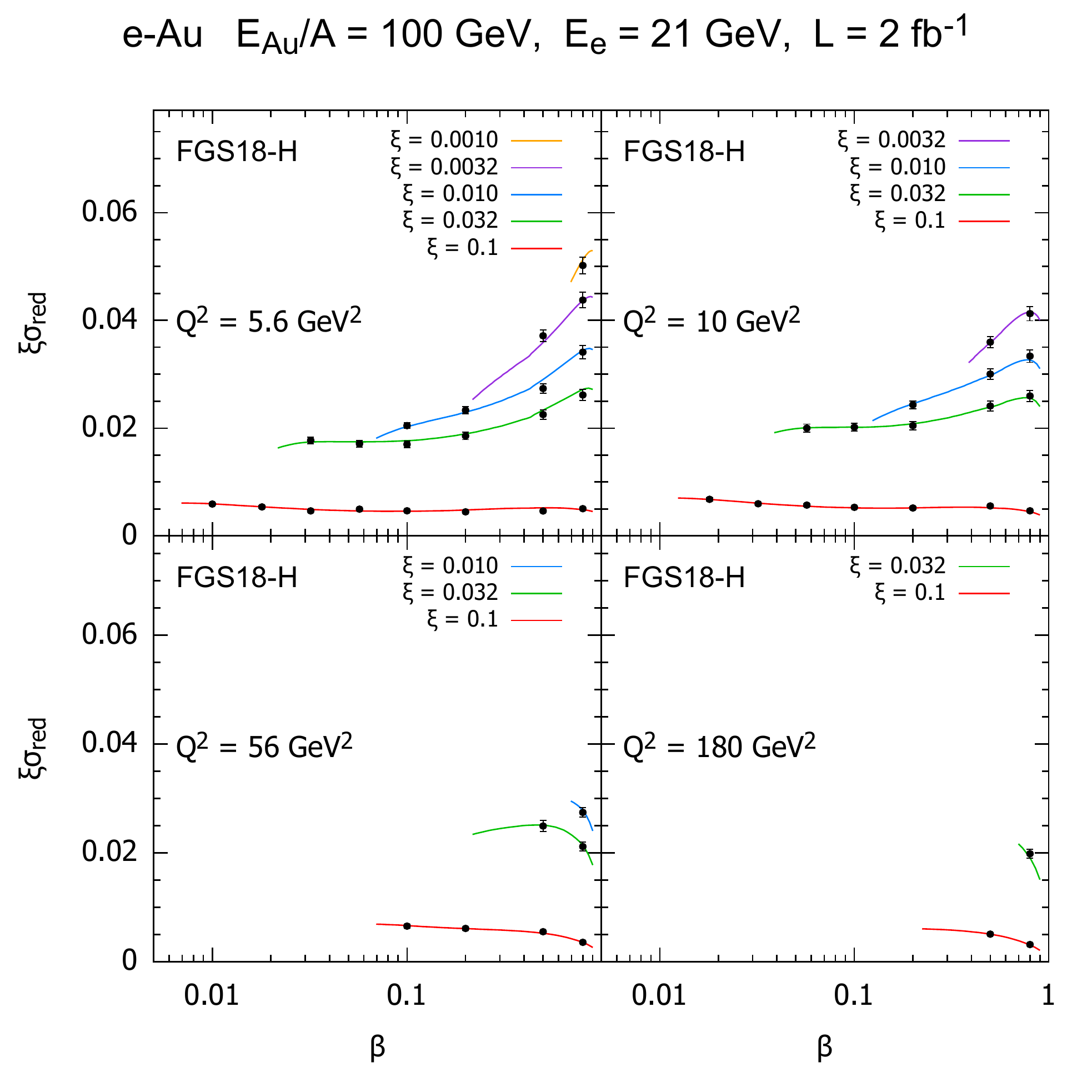}%
}
\caption{Simulated data for the diffractive reduced cross section as a function of $\beta$ in bins of $\xi$ and $Q^2$ for $e\,^{197}$Au collisions at the EIC, in the models in \cite{Frankfurt:2011cs}.
The curves for $\xi = 0.032, 0.01, 0.0032, 0.001$ are shifted up by 0.005, 0.01, 0.015, 0.02, respectively.}
\label{fig:sigred_Au01e21}
\end{figure}

Pseudodata were generated using the same method, 5\% 
uncorrelated systematic error and luminosity 2 fb$^{-1}$ as 
described for $ep$ in Section \ref{sec:pseudo_data}. 
The results for the LHeC and FCC-eh are shown in Figs. \ref{fig:sigred_Pb7e60} 
and \ref{fig:sigred_Pb50e60}, respectively (for a selected subset of bins). 
The similarly large coverage and small uncertainty (dominated by the assumed systematics) 
illustrated in these two figures compared to Figs. \ref{fig:sigred_ep_lhec} and \ref{fig:sigred_ep_fcc} make it clear that an accurate extraction of nDPDFs in $^{208}$Pb in an extended kinematic region, similar to that shown in Figs. \ref{fig:pdf_fits_lhec}, \ref{fig:pdf_fits_fcc} and \ref{fig:pdf_7_50_xi}, will be possible.
We also include in Fig. \ref{fig:sigred_Au01e21} the corresponding results for $e$Au collisions at the EIC. Studies performed for $ep$ at those energies show that the expected accuracy for the extraction of DPDFs at the EIC is comparable to that in existing DPDFs for the proton at HERA. Assuming, as we did for the LHeC and FCC-eh, a similar experimental uncertainty, integrated luminosity and kinematic coverage, the accuracy in the extraction of nDPDFs at the EIC would then be similar to that of existing HERA fits.

\section{Conclusions}
\label{sec:conclusions}

In this paper we have investigated the potential 
of the LHeC and FCC-eh machines for the measurement of 
diffractive cross sections and to constrain 
the diffractive parton densities. The LHeC machine would extend the available kinematic range in $x$ by a factor of order $20$ and the maximum $Q^2$ by 
a factor of order $100$. The FCC-eh machine would extend 
the accessible region by an order of magnitude with respect to 
LHeC both in $x$ and $Q^2$.  This translates into a range of available 
$\xi$ down to $10^{-4}$ at the LHeC and down to $10^{-5}$ 
for FCC-eh for a wide range of $\beta$. 
With the assumed very conservative integrated luminosity of $2 \,{\rm fb}^{-1}$ we have  generated large pseudodata sets of $1200- 1800$ points for the 
LHeC and of $1700-2600$ points for the FCC-eh, depending on the minimum $Q^2$. 
The simulated data  have very small error bars,  dominated by the assumed $5\%$ systematic error. We have performed fits of the diffractive parton densities to the simulated pseudodata, following the methodology employed previously at HERA. The DPDF determination using the 
pseudodata substantially improves the precision achieved in the HERA analysis, reducing the DPDF uncertainties by a factor  $5-7$
for the LHeC and $10-15$ for the FCC-eh. 

We stress that the uncertainty bands shown in the corresponding plots come purely from experimental errors. No attempt is made to 
evaluate theoretical sources of uncertainty, due for example to fixed parameters in the initial conditions or the evolution
or to the functional form of the parton parametrization at the starting scale.
This corresponds to our aim of establishing the experimental precision achievable in these new machines. Besides, 
if the luminosity were increased one could  perform a finer binning and constrain the extracted DPDFs even more.

The accuracy of the DPDF extraction depends only mildly on  the 
maximal value of $\xi$. In particular, we found that changing  
$\xi$ from $0.32$ to $0.1$ has a negligible impact on the 
precision of the extracted DPDFs.  This is very encouraging since the 
large $\xi$ region is very challenging experimentally and theoretically.  
On the other hand, we found a rather large sensitivity to the  
functional form of the gluon DPDF; specifically, a flat and non-flat gluon -- which were indistinguishable at HERA -- produce sizeably different $\chi^2/{\rm ndf}$ at the LHeC and FCC-eh. Besides, the fits are also sensitive to the assumed minimal value of $Q^2$ 
used in the DGLAP fits. This feature is understandable since the 
DGLAP evolution is very sensitive to the low $Q^2$ region, which is 
crucial for constraining the initial condition. This fact indicates the potential of both machines to constrain the diffractive parton densities in this region and, eventually, physics that goes beyond the standard twist-2 DGLAP evolution.  Finally, we have investigated the possibility of inclusive diffraction in the case of nuclear targets. Using models which employ Gribov inelastic shadowing, we make predictions for the nuclear ratios for the diffractive structure functions $F_2$ and $F_L$, and provide the simulated data sets.  
We find that the accurate measurement of the nuclear diffractive cross section 
would be possible in the nuclear case, with similar 
coverage in $\beta,\xi$ and $Q^2$ and similar precision to the proton case. 

The extended kinematic range of both machines offers new exciting possibilities in 
diffraction. One is that they are sensitive to the top contribution 
to diffraction. Since HERA did not give access to the top, 
none of the models used to simulate the pseudodata provides 
a reliable contribution from the top quark.  In the 
present analysis the top contribution was thus neglected, 
but it could be investigated in  further studies, particularly for the FCC-eh. 
Furthermore, diffractive dijets could also be included and their impact on the extraction of DPDFs evaluated.
Another interesting possibility is that of charged current diffraction. This was measured at HERA but in a very limited kinematic range and with very small statistics. In  future DIS machines this would certainly be a much better explored process and 
would provide additional tests for factorization in diffraction. 

Summarizing, both the LHeC and its higher energy version, 
the FCC-eh, offer unprecedented capabilities for studying 
diffraction both in $ep$ and  $eA$. This first exploratory study 
illustrates some of the  huge range of opportunities. More extensive studies, both on the phenomenological side and at detector level, are left for the future. These new possibilities for investigating  proton and nuclear structure will eventually open new avenues in the understanding of dynamics beyond linear evolution, such as 
higher twists and non-linear effects, and, ultimately, hopefully, confinement.

\section*{Acknowledgements}

We thank Vadim Guzey for providing  the FGS parametrization and Max Klein for reading the manuscript. We also thank  John Collins and Krzysztof Golec-Biernat for discussions. NA was supported by Ministerio de Ciencia e
Innovaci\'on of Spain under projects FPA2014-58293-C2-1-P, FPA2017-83814-P and Unidad de Excelencia Mar\'{\i}a de
Maetzu under project MDM-2016-0692, by Xunta de Galicia (Conseller\'{\i}a de Educaci\'on) within the Strategic Unit
AGRUP2015/11, and by FEDER. This work has been performed
in the framework of COST Action CA15213 `Theory of hot matter and relativistic heavy-ion collisions' (THOR). 
WS was supported by the National Science centre, Poland, Grant No. 2014/13/B/ST2/02486.
AMS was supported by the  Department of Energy Grant No. DE-SC-0002145,
 as well as the National Science centre, Poland, Grant No. 2015/17/B/ST2/01838.



\end{document}